\documentclass[twocolumn,11pt]{IEEEtran}
\usepackage[cmex10]{amsmath}
\usepackage{amsfonts}
\usepackage{color,graphicx}
\usepackage[bookmarks=true]{hyperref}
\usepackage{subfig}
\usepackage{latexsym}
\usepackage{amssymb}
\usepackage{nicefrac}
\usepackage{xspace}
\usepackage{paralist}
\usepackage{xifthen}% provides \ifthenelse and \isempty
\usepackage{tikz}
\usetikzlibrary{shapes,backgrounds}

\newtheorem{theorem}{Theorem}[section]  
\newtheorem{corollary}[theorem]{Corollary}
\newtheorem{lemma}[theorem]{Lemma}
\newtheorem{prop}[theorem]{Proposition}
\newtheorem{definition}{Definition}[section]  
\newtheorem{eg}{Example}[section]  

\newenvironment{thm}{\medskip\begin{theorem}}{\end{theorem}\medskip}
\newenvironment{corol}{\medskip\begin{corollary}}{\end{corollary}\medskip}
\newenvironment{lem}{\medskip\begin{lemma}}{\end{lemma}\medskip}
\newenvironment{defn}{\medskip\begin{definition}}{\end{definition}\medskip}

\newcounter{MYtempeqncounter}
\newcounter{MYtempeqncounter1}

\hypersetup{
   colorlinks=true,
   pdftitle={},
   pdfauthor={Prabhakaran \& Prabhakaran},
   citecolor=black,
   linkcolor=black
}
\newcommand{\namedref}[2]{\hyperref[#2]{#1~\ref*{#2}}}
\newcommand{\sectionref}[1]{\namedref{Section}{sec:#1}}
\newcommand{\theoremref}[1]{\namedref{Theorem}{thm:#1}}
\newcommand{\definitionref}[1]{\namedref{Definition}{def:#1}}
\newcommand{\corollaryref}[1]{\namedref{Corollary}{cor:#1}}
\newcommand{\exampleref}[1]{\namedref{Example}{eg:#1}}
\newcommand{\lemmaref}[1]{\namedref{Lemma}{lem:#1}}

\newcommand{\figureref}[1]{\namedref{Figure}{fig:#1}}
\newcommand{\appendixref}[1]{\namedref{Appendix}{app:#1}}

\providecommand*\Exp[2][]{%
\ifthenelse{\isempty{#1}}{%
\ensuremath{{\mathbb E}\left[#2\right]}}{%
\ensuremath{{\mathbb E}_{#1}\left[#2\right]}}%
}
\providecommand*\typical[2][]{%
\ifthenelse{\isempty{#1}}{%
\ensuremath{{\mathcal T}\upn_{#2}}}{%
\ensuremath{{\mathcal T}\upn_{#2}(#1)}}%
}

\newcommand{\phast}{{\ensuremath{\vphantom{\ast}}}}

\newcommand{\defineqq}{\triangleq}
\newcommand{\namedeqq}[1]{\ensuremath{\stackrel{\mathrm{(#1)}}{=}}}
\newcommand{\namedleq}[1]{\ensuremath{\stackrel{\mathrm{(#1)}}{\le}}}
\newcommand{\namedgeq}[1]{\ensuremath{\stackrel{\mathrm{(#1)}}{\ge}}}

\newcommand{\Real}{\ensuremath{\mathbb{R}}\xspace}
\newcommand{\Rplus}{\ensuremath{{\mathbb{R}}_+}\xspace}

\newcommand{\mchain}{\ensuremath{-}}

\newcommand{\Com}[1]{\ensuremath{{\mathrm C}_\text{\sf #1}}\xspace}
\newcommand{\GWlower}{\ensuremath{\mathcal L}_\text{\sf GW}\xspace}

\newcommand{\ihull}[1]{\ensuremath{i\left(#1\right)}}

\newcommand{\A}{\ensuremath{\text{\sf A}}\xspace}
\newcommand{\B}{\ensuremath{\text{\sf B}}\xspace}
\newcommand{\C}{\ensuremath{\text{\sf C}}\xspace}
\newcommand{\CI}{\ensuremath{\text{\sf CI}}\xspace}
\newcommand{\RI}{\ensuremath{\text{\sf RI}}\xspace}

\newcommand{\sA}{\ensuremath{\mathcal A}\xspace}
\newcommand{\sB}{\ensuremath{\mathcal B}\xspace}
\newcommand{\sC}{\ensuremath{\mathcal C}\xspace}
\newcommand{\sE}{\ensuremath{\mathcal E}\xspace}
\newcommand{\sQ}{\ensuremath{\mathcal Q}\xspace}
\newcommand{\sS}{\ensuremath{\mathcal S}\xspace}
\newcommand{\sX}{\ensuremath{\mathcal X}\xspace}
\newcommand{\sY}{\ensuremath{\mathcal Y}\xspace}
\newcommand{\sU}{\ensuremath{\mathcal U}\xspace}
\newcommand{\sV}{\ensuremath{\mathcal V}\xspace}
\newcommand{\shS}{\ensuremath{\widehat{\mathcal S}}\xspace}
\newcommand{\hL}{\ensuremath{\widehat{L}\xspace}}
\newcommand{\hS}{\ensuremath{\widehat{S}\xspace}}
\newcommand{\hX}{\ensuremath{\widehat{X}\xspace}}
\newcommand{\hY}{\ensuremath{\widehat{Y}\xspace}}
\newcommand{\hl}{\ensuremath{\hat{l}\xspace}}
\newcommand{\hs}{\ensuremath{\hat{s}\xspace}}
\newcommand{\tl}{\ensuremath{\tilde{l}\xspace}}
\newcommand{\ts}{\ensuremath{\tilde{s}\xspace}}
\newcommand{\bbZ}{\ensuremath{\mathbb Z}\xspace}

\newcommand{\sR}{\ensuremath{\mathcal R}\xspace}

\newcommand{\Del}{\ensuremath{\widehat\phi}\xspace}
\newcommand{\del}{\ensuremath{\widehat\delta}\xspace}

\newcommand{\aci}{{ACI}\xspace}
\newcommand{\gw}{{GW}\xspace}
\newcommand{\gk}{{GK}\xspace}

\newcommand{\Wyner}{\ensuremath{\sf Wyner}}
\newcommand{\ACI}{\ensuremath{\text{\sf ACI}}}
\newcommand{\ARI}{\ensuremath{\text{\sf ARI}}}
\newcommand{\GW}{\ensuremath{\text{\sf GW}}\xspace}
\newcommand{\GKW}{\ensuremath{\text{\sf GK}}\xspace}
\newcommand{\sRsACI}{\ensuremath{{\mathcal R}_\ACI}}
\newcommand{\sRsGW}{\ensuremath{{\mathcal R}_\GW}}
\newcommand{\sRsGKW}{\ensuremath{{\mathcal R}_\ARI}}

\newcommand{\sP}{\ensuremath{{\mathcal P}}\xspace}
\newcommand{\sPXY}{\ensuremath{\sP_{X,Y}}\xspace}
\newcommand{\sPXUVY}{\ensuremath{\sP_{XUVY}}\xspace}
\newcommand{\hsPXY}{\ensuremath{\widehat\sP_{X,Y}}\xspace}
\newcommand{\pQXY}{\ensuremath{{p_{Q|XY}}}\xspace}
\newcommand{\pQXUVY}{\ensuremath{{p_{Q|XUVY}}}\xspace}

\newcommand{\tens}[3]{\ensuremath{T({#1};{#2}|{#3})}\xspace}
\newcommand{\RT}{\ensuremath{\mathfrak{T}}\xspace}
\newcommand{\Rtens}[2]{\ensuremath{\RT({#1};{#2})}\xspace}
\newcommand{\Tintx}[2]{\ensuremath{T^{\mathrm{int}}_{1}({#1};{#2})}\xspace}
\newcommand{\Tinty}[2]{\ensuremath{T^{\mathrm{int}}_{2}({#1};{#2})}\xspace}
\newcommand{\Tintz}[2]{\ensuremath{T^{\mathrm{int}}_{3}({#1};{#2})}\xspace}

\newcommand{\RACIintz}[2]{\ensuremath{{\sRsACI}_{,3}^{\mathrm{int}}({#1};{#2})}\xspace}

\newcommand{\upn}{\ensuremath{^{(n)}}}

\newcommand{\Mfunc}{\ensuremath{\mathcal{M}}\xspace}
\newcommand{\M}[2]{\ensuremath{\Mfunc({#1};{#2})}\xspace}
\newcommand{\Kfunc}{\ensuremath{\RT}\xspace}
\newcommand{\KK}[2]{\ensuremath{\RT({#1};{#2})}\xspace}

\newcommand{\tp}{\ensuremath{\tilde{p}}}

\newcommand{\pt}{\ensuremath{\mathbf{a}}\xspace}

\newcommand{\pialice}{\ensuremath{\pi_{\mathrm{Alice}}}\xspace}
\newcommand{\pibob}{\ensuremath{\pi_{\mathrm{Bob}}}\xspace}
\newcommand{\Pialice}{\ensuremath{\Pi^{\mathrm{view}}_{\mathrm{Alice}}}\xspace}
\newcommand{\Pibob}{\ensuremath{\Pi^{\mathrm{view}}_{\mathrm{Bob}}}\xspace}
\newcommand{\Pialiceout}{\ensuremath{\Pi^{\mathrm{out}}_{\mathrm{Alice}}}\xspace}
\newcommand{\Pibobout}{\ensuremath{\Pi^{\mathrm{out}}_{\mathrm{Bob}}}\xspace}

\newcommand{\simPialice}{\ensuremath{\Sigma^{\mathrm{view}}_{\mathrm{Alice}}}\xspace}
\newcommand{\simPibob}{\ensuremath{\Sigma^{\mathrm{view}}_{\mathrm{Bob}}}\xspace}

\newcommand{\samples}[3]{\ensuremath{{#1}^{#2}\leadsto{#3}}\xspace}
\newcommand{\statsamples}[4]{\ensuremath{{#2}^{#3}\overset{#1}{\leadsto}{#4}}\xspace}

\usepackage[T1]{fontenc}

\begin{document}

\title{Assisted Common Information with an Application to Secure Two-Party Sampling}

\author{Vinod M. Prabhakaran and Manoj M. Prabhakaran
\thanks{This work was presented in part at IEEE International Symposia on
Information Theory (ISIT) 2010 and
2011~\cite{PrabhakaranPr10,PrabhakaranPr11}.
Vinod M. Prabhakaran's work was supported in part by a Ramanujan Fellowship from
the Department of Science and Technology, Government of India. Manoj M. Prabhakaran's work was supported in part by NSF CAREER award 07-47027 and NSF grant 12-28856.

Vinod M. Prabhakaran is with the School of Technology and Computer Science, Tata Institute of Fundamental Research, Mumbai 400 005 India (email: vinodmp@tifr.res.in)

Manoj M. Prabhakaran is with the Department of Computer Science, University of Illinois, Urbana-Champaign, IL 61801 USA (email: mmp@illinois.edu)
}%
}

\maketitle

\begin{abstract}
An important subclass of secure multiparty computation is secure sampling:
two parties output samples of a pair of jointly distributed random
variables such that neither party learns more about the other party's
output than what its own output reveals. The parties make use of a setup
--- correlated random variables with a different distribution --- as well
as unlimited noiseless communication. An upperbound on the rate of producing samples of a desired distribution from a given setup is presented.

The region of {\em tension} developed in this paper measures how well the
dependence between a pair of random variables can be resolved by a piece of
common information. The bounds on rate are a consequence of a monotonicity property: a protocol between two parties can only lower the tension between their ``views''.

Connections are drawn between the region of tension and the notion of common information. A generalization of the G\'acs-K\"orner common information, called the Assisted Common Information, which takes into account ``almost common'' information ignored by G\'acs-K\"orner common information is defined. The region of tension is shown to be related to the rate regions of both the Assisted Common Information and the Gray-Wyner systems (and, a fortiori, Wyner's common information).
\end{abstract}

\section{Introduction}

Secure multi-party computation is a central problem in modern cryptography.
Roughly, the goal of secure multi-party computation is to carry out
computations on inputs distributed among two (or more) parties, so as to
provide each of them with no more information than what their respective inputs
and outputs reveal to them. Our focus in this paper is on an important
sub-class of such problems --- which we shall call {\em secure 2-party
sampling} --- in which the computation has no inputs, but the outputs to the
parties are required to be from a given joint distribution (and each party
should not learn anything more than its part of the output). Also we shall
restrict ourselves to the case of honest-but-curious adversaries.  It is
well-known (see, for instance, \cite{Wullschleger08thesis} and references
therein) that very few distributions can be sampled from in this way, unless
the computation is aided by a {\em set up} --- some jointly distributed random variables
that are given to the parties at the beginning of the protocol.  The set up
itself will be from some distribution $(X,Y)$ (Alice gets $X$ and Bob gets $Y$)
which is different from the desired distribution $(U,V)$ (Alice getting $U$ and
Bob getting $V$).  The fundamental question then is, which set ups $(X,Y)$ can
be used to securely sample which distributions $(U,V)$, and {\em at what
rate} (i.e., how many samples of $(U,V)$ can be generated per sample of
$(X,Y)$ used).

While the feasibility question can be answered using combinatorial analysis
(as, for instance, was done in \cite{Kilian00}), information theoretic
tools have been put to good use to show bounds on rate of protocols
(e.g.
\cite{Beaver96,DodisMi99,WinterNaIm03,ImaiMuNaWi04,ImaiMoNa06,CsiszarAh07,ImaiMoNa07,WolfWu08,WinklerWu10}).
Our work continues on this vein of using information theory to formulate
and answer rate questions in cryptography. Specifically, we
generalize the concept of common information~\cite{GacsKo73} as defined by
G\'acs and K\"orner~(GK) and use this generalization to establish upper
bounds on the rate of secure sampling.

Finding a meaningful definition for the ``common information'' of a pair of
dependent random variables $X$ and $Y$ has received much attention starting
from the
1970s~\cite{GacsKo73,Witsenhausen75,Wyner75,AhlswedeKo74,Yamamoto94}.  We
propose a new measure --- a three-dimensional region --- which brings out a
detailed picture of the extent of common information of a pair.  This gives
us an expressive means to compare different pairs with each other, based on
the shape and size of their respective regions. Besides the specific
application to secure sampling discussed in this paper, we believe that our
generalization may have potential applications in information theory,
cryptography, communication complexity (and hence complexity in various
computational models), game theory, and distributed control, where the role
of dependent random variables and common randomness is well-recognized.

Suppose $X=(X',Q)$ and $Y=(Y',Q)$ where $X',Y',Q$ are independent. Then a
natural measure of ``common information'' of $X$ and $Y$ is $H(Q)$.
$Q$ is determined both by $X$ and by $Y$, and further, conditioned on $Q$,
there is no ``residual information'' that correlates $X$ and $Y$ i.e.,
$X-Q-Y$.  One could extend this to arbitrary $X,Y$, in a couple of natural
ways. One approach, which corresponds to a definition of G\'{a}cs and
K\"{o}rner~\cite{GacsKo73}%
\footnote{This is not the {\em definition} of common information in
\cite{GacsKo73}, but the consequence of a non-trivial result in that work.
The original definition, which is in terms of a communication problem, is
detailed in \sectionref{ACI} (along with our extensions).}
is to find the ``largest'' random variable $Q$ that is determined by
$X$ alone as well as by $Y$ alone (with probability 1):
\begin{align}
&\Com\GKW(X;Y) = \max_{\substack{\pQXY:\\H(Q|X)=H(Q|Y)=0}} H(Q)\notag\\
&= I(X;Y) - \min_{\substack{\pQXY:\\H(Q|X)=H(Q|Y)=0}}
I(X;Y|Q).\label{eq:introGKCI}
\end{align}
Note that in this case, the common information is necessarily no more than
the mutual information, and in general this gap is non-zero, i.e., common
information, in general, does not account for all the dependence between
$X$ and $Y$. An alternate generalization, which corresponds to the approach
of Wyner~\cite{Wyner75}%
\footnote{Again, the actual definition of~\cite{Wyner75}, which is in terms
of a source coding problem, is different. The expression above is a
consequence of a result in~\cite{Wyner75}. The definition and results in
\cite{Wyner75} are described in \sectionref{GrayWyner}.},
is to consider the ``smallest'' random variable $Q$ so
that conditioned on $Q$ there is no residual mutual information.
Smallness of $Q$, in this case, is measured in terms of $I(XY;Q)$.
\begin{align}
&\Com\Wyner(X;Y) = \min_{\substack{\pQXY:\\X-Q-Y}} I(XY;Q) \notag\\
&= I(X;Y) + \min_{\substack{\pQXY:\\X-Q-Y}} (I(Y;Q|X)+I(X;Q|Y)). \label{eq:introWynCI}
\end{align}
Note that in this case, the common information is necessarily no less than
the mutual information.  
When $X,Y$
are of the form $X=(X',Q)$ and $Y=(Y',Q)$, where $X',Y',Q$ are independent,
then there indeed is a unique interpretation of common information (when
$\Com\GKW(X;Y)=\Com\Wyner(X;Y)=H(Q)$). But otherwise, between the
extremes represented by these two measures, there are several ways
in which one could define a random variable to capture the dependence 
between $X$ and $Y$. 

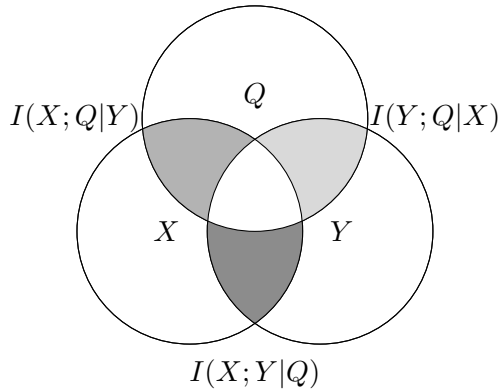
\begin{figure}[htb]
\centering

\def\firstcircle{(210:1cm) circle (1.5cm)}
\def\secondcircle{(90:1cm) circle (1.5cm)}
\def\thirdcircle{(330:1cm) circle (1.5cm)}

\begin{tikzpicture}
    \draw \firstcircle node[left] {$X$};
    \draw \secondcircle node [above] {$Q$};
    \draw \thirdcircle node [right] {$Y$};

    \begin{scope}
      \clip \firstcircle;
      \fill[gray!60] \secondcircle;
    \end{scope}
    \begin{scope}
      \clip \thirdcircle;
      \fill[gray!30] \secondcircle;
    \end{scope}
    \begin{scope}
      \clip \firstcircle;
      \fill[gray!90] \thirdcircle;
    \end{scope}

    \begin{scope}
      \clip \firstcircle;
      \clip \secondcircle;
      \fill[white] \thirdcircle;
    \end{scope}

\node at (23:2.6cm) {$I(Y;Q|X)$};
\node at (157:2.6cm) {$I(X;Q|Y)$};
\node at (270:2.4cm) {$I(X;Y|Q)$};

    \draw \firstcircle; 
    \draw \secondcircle; 
    \draw \thirdcircle;

\end{tikzpicture}

\caption{A Venn diagram representation of the three coordinates of $\tens XYQ$.}
\label{fig:venn}
\end{figure}

One way to look at the new quantities we introduce is as a way to capture an
entire spectrum of random variables that approximately capture the
dependence between $X$ and $Y$.  In \sectionref{tension} we shall define a
three-dimensional ``region of tension'' for $X,Y$, which measures how well
can the dependence between $X,Y$ be captured by a random variable.  In
\figureref{Rtens}, we schematically depict this region. Looking ahead, we
mark the quantities $I(X;Y)-\Com\GKW(X;Y)$ and $\Com\Wyner(X;Y)-I(X;Y)$
in this figure to illustrate the gap between mutual information and
the two notions of common information in terms of the region of tension. The boundary of the region of tension is made up of triples of the form $(I(Y;Q|X),I(X;Q|Y),I(X;Y|Q))$; see \figureref{venn}. G\'acs-K\"orner \eqref{eq:introGKCI} considers $Q$ for which the first two coordinates are 0, and Wyner's common information \eqref{eq:introWynCI} considers $Q$ for which the last coordinate is 0.

In \sectionref{ACI}, we give an operational meaning to the region of tension by
generalizing the setting of G\'{a}cs-K\"{o}rner (see \figureref{common}) to the ``Assisted
Common Information system.'' We show that the associated rate regions are
closely related to the region of tension (\corollaryref{GKW-tension}). In \sectionref{GrayWyner}, we
consider the Gray-Wyner system~\cite{GrayWy74} (which can be viewed as a generalization of \Com\Wyner)
and show that the rate region associated with this system is also closely
related to the region of tension (\theoremref{affine}).  This clarifies the connection between
\Com\GKW and the Gray-Wyner system. In particular, previously known
connections readily follow from our results. Further, we show how two
quantities identified in recent work in the context of lossless coding with
side-information~\cite{MarcoEf09} and the Gray-Wyner
system~\cite{KamathAn10} can be obtained in terms of the region of tension
(\corollaryref{cornerconnection}).

Quite apart from the information theoretic questions related to common
information, our motivating application for defining the region of tension
is the cryptographic problem of bounding the rate of secure-sampling
described at the beginning of this article.  In \sectionref{crypto}, we show that the region of tension
of the views of two parties engaged in such a protocol can only
monotonically lower (expand towards the origin) and not rise (shrink away
from the origin).  Thus, by comparing the regions for the target random
variables and the given random variables, we obtain improved upperbounds on
the rate at which one pair can be
securely generated using another.
This bound is stated in \corollaryref{tension-stat-rate-bound}.

We also illustrate an interesting example (in \sectionref{example}) where we
obtain a tight upperbound, strictly improving on the prior work.
This example considers the rate at which random samples of ``(bit) oblivious
transfer'' (OT) --- an important cryptographic primitive --- can be securely
generated from a variant of it. The latter variant consists of two ``{\em
string} oblivious transfer'' (string OT) instances, {\em one in each
direction}.  Intuitively, this variant is quantitatively much more complex
than bit oblivious transfer, and the complexity increases with the length of
the strings involved. Prior bounds leave open the possibility that by using
longer strings in string OT, one can increase the rate at which bit OT
instances can be securely sampled per instance of string OT used.  But by
comparing the regions of tension, we can show that this is not the case: we
show that using arbitrarily long strings in the string OT yields the same
rate as using strings that are a single bit long!

\paragraph*{Outline} \sectionref{tension} defines the region of tension for a
pair of correlated random variables, and establishes some of its properties.
\sectionref{ACI} and \sectionref{GrayWyner} introduce the concepts of common
information \Com\GKW and \Com\Wyner in terms of the G\'acs-K\"orner and
Gray-Wyner systems (and a new generalization, in the case of the former), and
establish the connections with the region of tension. \sectionref{crypto}
defines the secure sampling problem, a monotonicity property of the region of
tension and its application in bounding the rate of secure sampling.  The
reader may choose to read only \sectionref{tension}, \sectionref{ACI} and
\sectionref{GrayWyner} for the results on common information, or alternatively
only \sectionref{tension} and \sectionref{crypto} for results on secure
two-party sampling.

\section{Tension and the Region of Tension}
\label{sec:tension}

Now we introduce our main tool which generalizes \gk common
information and also serves as a measure of cryptographic complexity
of securely sampling a pair of random variables. Intuitively, we measure how
well common information captures (or does not capture) the mutual information between a pair
of random variables $(X,Y)$.

\subsection{Definitions}

Throughout this paper we concern ourselves with pairs of correlated {\em
finite} random variables $(X,Y)$ with joint distribution (p.m.f.)
$p_{X,Y}$. \sX and \sY shall stand for the (finite) alphabets of $X$ and
$Y$ respectively.  We let \sPXY denote the set of all random variables $Q$
jointly distributed with $(X,Y)$ --- i.e., all conditional p.m.f.s
$p_{Q|X,Y}$.

The {\em total variation distance}%
\footnote{In cryptography literature, $\Delta(\cdot,\cdot)$ is more commonly
called statistical difference.}
between two random variables $X$ and $X'$ over the same alphabet $\sX$ 
is $\Delta(X,X') \defineqq \frac12 || p_{X} - p_{X} ||_1 
= \frac12 \sum_{x\in\sX} |p_{X}(x)-p_{X'}(x)|$.
$H_2(.)$ will denote the binary entropy function: $H_2(p)\defineqq
p\log(1/p)+(1-p)\log(1/(1-p))$ (for $0< p < 1$), and $H_2(0)=H_2(1)=0$. All logarithms will be to the base 2.

The {\em characteristic bipartite graph} of a pair of correlated random
variables $(X,Y)$ is the graph with vertices in $\sX\cup\sY$ and an edge
between $x\in\sX$ and $y\in\sY$ if and only if $p_{XY}(x,y)>0$. (See
\figureref{zsource} for an example.)

Now we give the main definitions of this section.

\begin{defn}
\label{def:resolve}
For a pair of correlated random variables $(X,Y)$, and
$\pQXY\in\sPXY$, we say $Q$ {\em perfectly resolves} $(X,Y)$ if $I(X;Y|Q)=0$ and
$H(Q|X)=H(Q|Y)=0$. We say $(X,Y)$ is {\em perfectly resolvable}
if there exists $\pQXY\in\sPXY$ such that $Q$ perfectly resolves $(X,Y)$.
\end{defn}
If $(X,Y)$ is perfectly resolvable, then their \gk common information 
represents the entire mutual information between them, i.e., \gk common information is equal to the mutual information (see \eqref{eq:introGKCI}).  We intend to measure the extent to which a given $(X,Y)$ is
{\em not} perfectly resolvable. Towards this we introduce a 3-dimensional
measure called {\em tension} of $(X,Y)$, defined as follows.

\begin{defn}
\label{def:tension}
For a pair of correlated random variables $(X,Y)$ and $\pQXY\in\sPXY$,
the {\em tension} of $(X,Y)$ given $Q$ is denoted by $\tens XYQ \in \Rplus^3$
and defined as $\tens XYQ \defineqq \big(I(Y;Q|X),I(X;Q|Y),I(X;Y|Q)\big)$.
The {\em region of tension of $(X,Y)$}, denoted by $\Rtens XY \subseteq \Rplus^3$ is defined as
\[ \Rtens XY \defineqq \ihull{\{ \tens XYQ: \pQXY \in \sPXY\}}, \]
where $\ihull{{\sf S}}$ denotes the {\em increasing hull} of ${\sf S}\subseteq
\Rplus^3$, defined as $\ihull{\sf S}\defineqq\{s\in\Rplus^3:  \exists s'\in {\sf S} \text{ s.t. } s\geq s' \}$.%
\footnote{For two vectors $(x,y,z),(x',y',z')\in\Rplus^3$, we write
$(x,y,z)\ge(x',y',z')$ to mean $x\ge x'$, $y\ge y'$ and $z\ge z'$.}
\end{defn}
Since we consider only random variables with finite alphabets \sX and \sY,
it follows from Fenchel-Eggleston's strengthening of Carath\'{e}odory's
theorem~\cite[pg. 310]{CsiszarKo81}, that we can restrict ourselves to
$\pQXY\in\sPXY$ with alphabet \sQ such that $|\sQ|\leq |\sX||\sY|+2$. More
precisely,
\begin{align}
\label{eq:Rtens-cardbd}
\Rtens XY = \ihull{\{ \tens XYQ: \pQXY \in \hsPXY\}},
\end{align}
where
\hsPXY is defined as the set of all conditional p.m.f.'s $p_{Q|X,Y}$ such
that the cardinality of alphabet $\sQ$ of $Q$ is such that $|\sQ|\leq
|\sX||\sY|+2$.

\begin{figure}[htb]
\centering
\scalebox{0.6}{\includegraphics{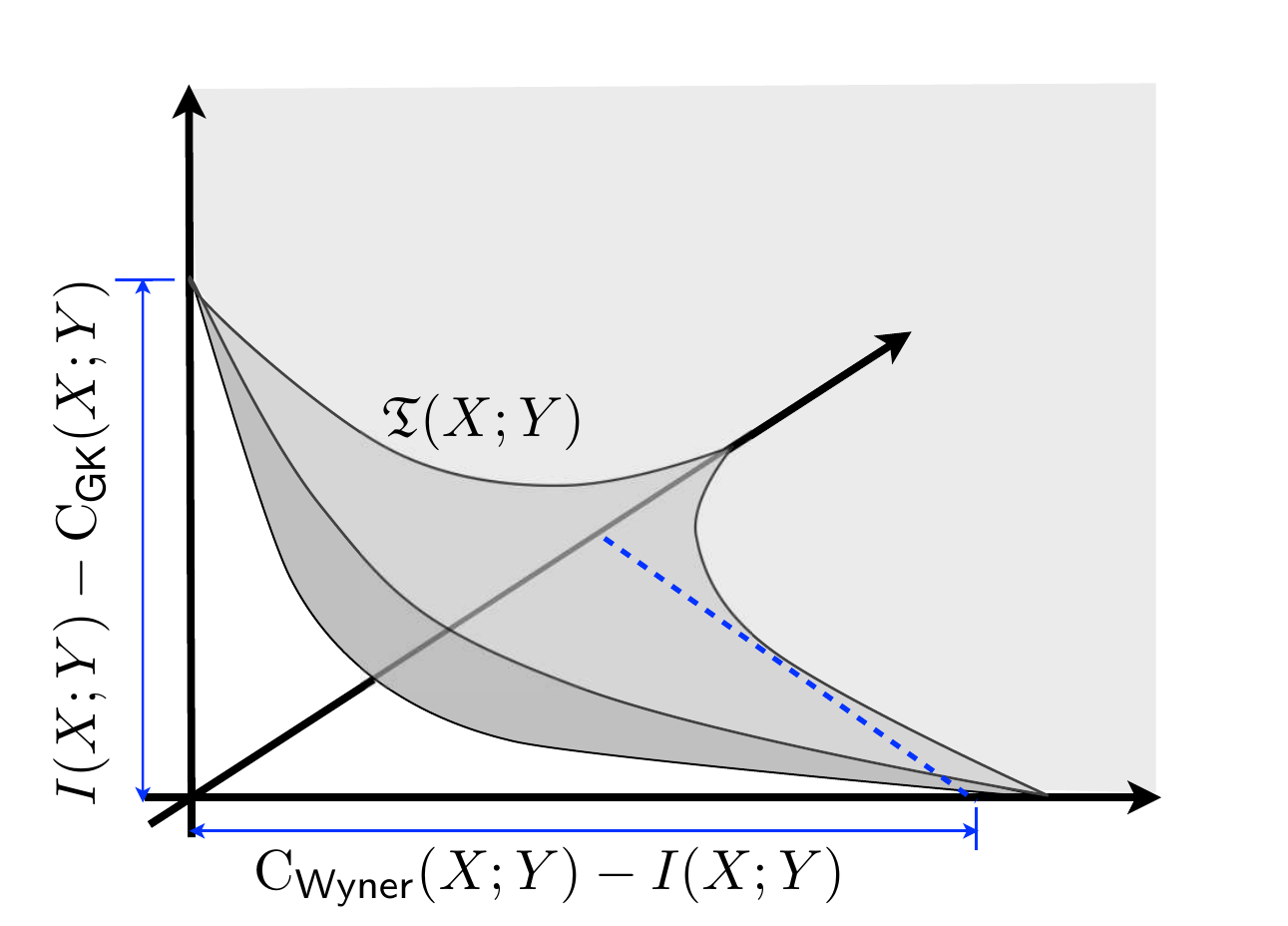}}
\caption{A schematic representation of the region \Rtens XY.
\Rtens XY is an unbounded, convex region, bounded away from the origin
(unless $(X,Y)$ is perfectly resolvable). Relationship between two
points on the boundary of \Rtens XY and the quantities
\Com\GKW(X;Y) and \Com\Wyner(X;Y) (see \eqref{eq:CGKWfromRIO} and
\eqref{eq:CWynerfromRtens}) is shown. (The dotted line is at 45$^\circ$
to the axes.)}
\label{fig:Rtens}
\end{figure}

We point out that \Rtens XY intersects all three axes (e.g., consider
$Q=Y$, $Q=X$ and $Q=0$, respectively). It will be of interest to consider
the three axes intercepts of the boundary of \Rtens XY.
\begin{align}
\begin{split}
\Tintx XY &\defineqq \min\{r_1:(r_1,0,0)\in\Rtens XY\}\\
\Tinty XY &\defineqq \min\{r_2:(0,r_2,0)\in\Rtens XY\}\\
\Tintz XY &\defineqq \min\{r_3:(0,0,r_3)\in\Rtens XY\}
\end{split}\label{eq:Tint}
\end{align}
The use of $\min$ instead of $\inf$ anticipates \theoremref{Rtens-closed}
which shows that $\Rtens XY$ is closed.

\subsection{Some Properties of Tension}

Firstly, we have an easy observation.
\begin{thm}
\label{thm:Rtens-resolvable}
\Rtens XY includes the origin 
if and only if the pair $(X,Y)$
is perfectly resolvable.
\end{thm}
\begin{IEEEproof}
We need to show that there exists \pQXY such that
$I(Y;Q|X)=I(X;Q|Y)=I(X;Y|Q)=0$ if and only if
there exists $p_{Q'|XY}$ such that
$H(Q'|X)=H(Q'|Y)=I(X;Y|Q')=0$. Clearly, the second
condition implies the first by taking $Q$ to be the
same as $Q'$. The converse follows from
\lemmaref{connectedcomp} which shows that
given \pQXY such that
$I(Y;Q|X)=I(X;Q|Y)=0$, we can find 
a random variable $Q'$ with 
$H(Q'|X)=H(Q'|Y)=0$ and $Q-Q'-XY$; 
 then, by \lemmaref{connectedcomp-addon}
it follows that $I(X;Y|Q') \le I(X;Y|Q)$, and 
hence $I(X;Y|Q)=0$ implies $I(X;Y|Q')=0$.
\end{IEEEproof}

The more interesting case is when \Rtens XY does not contain the origin,
and hence $(X,Y)$ is not perfectly resolvable.
Note that it is important to consider all three coordinates of
\tens XYQ together to identify the unresolvable nature of a pair $(X,Y)$,
because, as observed above, \Rtens XY does intersect each of the three
axes, or in other words, any two coordinates of \tens XYQ can be made
simultaneously 0 by choosing an appropriate $Q$.

As it turns out, the axes intercepts are identical to three quantities 
identified by Wolf and Wullschleger~\cite{WolfWu08}. In \cite{WolfWu08} 
these quantities were defined as 
\begin{align*}
H(X\searrow Y | Y) && H(Y\searrow X| X) && I(X;Y|X\wedge Y) 
\end{align*}
where, $X\searrow Y$ stands for the part of $X$ which depends on
$Y$ (i.e., a function of $X$ which distinguishes between different values of
$X$ if and only if they induce different conditional distributions on $Y$), and
$X\wedge Y$ stands for the {\em common information} between $X$ and $Y$
(i.e., the ''maximal'' function of $X$ that is also a function of $Y$, as
discussed in more detail in \sectionref{ACI}).
More precisely, the three quantities considered there are such that:
{\begin{align*}
H(Y\searrow X|X) &= \min_{\pQXY:H(Q|Y)=I(X;Y|Q)=0} H(Q|X) \\
H(X\searrow Y|Y) &= \min_{\pQXY:H(Q|X)=I(X;Y|Q)=0} H(Q|Y) \\
I(X;Y|X\wedge Y) &= \min_{\pQXY:H(Q|X)=H(Q|Y)=0} I(X;Y|Q).
\end{align*}}
In the appendix we prove the following theorem that these three quantities are the same
as $(\Tintx XY,\Tinty XY, \Tintz XY)$.
\begin{thm}
\label{thm:intercepts}
{\begin{align}
\Tintx XY &= \min_{\substack{\pQXY:\\H(Q|Y)=I(X;Y|Q)=0}} H(Q|X) \label{eq:Tintx}\\
\Tinty XY &= \min_{\substack{\pQXY:\\H(Q|X)=I(X;Y|Q)=0}} H(Q|Y) \label{eq:Tinty} \\
\Tintz XY &= \min_{\substack{\pQXY:\\H(Q|X)=H(Q|Y)=0}} I(X;Y|Q) \label{eq:Tintz}.
\end{align}}
\end{thm}

\paragraph*{Monotonicity of \Rtens XY} Wolf and Wullschleger showed that these three quantities have a certain
``monotonicity'' property (they can only decrease, as $X,Y$ evolve as the
views of two parties in a secure protocol).  We shall see that the
monotinicity of all the three quantities is a consequence of the
monotinicity of the entire region \Rtens XY. We define the precise nature of
this monotonicity in \sectionref{monotone} and prove it for \Rtens XY
in \sectionref{Rtens-monotone}.

The following result (proven in \appendixref{tension}) will be useful in
defining a ``multiplication'' operation on the region of tension as
a scaling (see \eqref{eq:convex-mult}). This in turn would be useful
in relating the region of tension and the rate of secure sampling, in
\sectionref{crypto}.
\begin{thm}
\label{thm:Rtens-convex}
The region \Rtens XY is convex.
\end{thm}

In extending the results in \sectionref{crypto} to statistical security
(rather than perfect security), the following results  would be important.
Firstly, the region of tension is closed.
\begin{thm}
\label{thm:Rtens-closed}
The region \Rtens XY is closed.
\end{thm}
\begin{IEEEproof}
By \eqref{eq:Rtens-cardbd}, and the fact that the increasing hull of a
compact set is closed (see \lemmaref{closedness-of-ihull} in
\appendixref{tension}), it is enough to show that $\{ \tens XYQ: \pQXY \in
\hsPXY\}$ is compact (i.e., closed and bounded (Heine-Borel theorem)).  For
this, notice that $\tens XYQ$ as a function of $p_{Q|XY}$ -- i.e., as a
function from \hsPXY to $\Real^3$ -- is continuous. Moreover, $\hsPXY$ is
compact. Since the image of a compact set under a continuous function is
compact,  $\{ \tens XYQ: \pQXY \in \hsPXY\}$ is compact. 
\end{IEEEproof}

Secondly,
the region of tension is {\em continuous} in the sense
that when the joint p.m.f.
$p_{X,Y}$ is close to the joint p.m.f. $p_{X',Y'}$, the tension
regions $\Rtens {X}{Y}$ and $\Rtens {X'}{Y'}$ are also close. We
measure closeness of these two joint p.m.f.'s (assumed without loss of
general to be defined over the same alphabet $\sX\times\sY$) by their total
variation distance $\Delta(XY,X'Y')$.

\begin{thm}
\label{thm:Rtens-continuity}
Suppose $\Delta(XY,X'Y')= \epsilon$, for some $\epsilon \in [0,1]$. Then,
$\Rtens {X}{Y} \subseteq \Rtens {X'}{Y'} - \delta(\epsilon)$,
where
$\delta(\epsilon) = 2H_2(\epsilon) + \epsilon\log\max\{|\sX|,|\sY|\}$,
 and for
${\sf S}\in\Real^3, \alpha\in\Real$, the notation ${\sf S}-\alpha$ stands for
$\{(r_1-\alpha,r_2-\alpha,r_3-\alpha):(r_1,r_2,r_3)\in {\sf S}\}$.
\end{thm}

\begin{IEEEproof}
Suppose $(r_1,r_2,r_3)\in \Rtens {X}{Y}$. We shall show that
$(r_1+\delta(\epsilon),r_2+\delta(\epsilon),r_3+\delta(\epsilon))\in \Rtens
{X'}{Y'}$. Since $(r_1,r_2,r_3)\in\Rtens {X}{Y}$, there is a
$p_{Q|X,Y}\in\sPXY$ such that $I(Y;Q|X)\leq r_1$, $I(X;Q|Y)\leq
r_2$, and $I(X;Y|Q)\leq r_3$. Let $p_{Q'|X',Y'}=p_{Q|X,Y}$.
It is enough to prove that 
\begin{align*}
I(Y';Q'|X') &\leq I(Y;Q|X) + \delta(\epsilon),\\
I(X';Q'|Y') &\leq I(X;Q|Y) + \delta(\epsilon),\\
I(X';Y'|Q') &\leq I(X;Y|Q) + \delta(\epsilon).
\end{align*}

We will make use of the following lemma which is proved in
\appendixref{tension}.
\begin{lem}
\label{lem:statdiff-mutualinfo}
Suppose random variables $(A,B,C)$ and $(A',B',C')$ over the same alphabet
$\sA\times\sB\times\sC$
are such that $\Delta(ABC,A'B'C')=\epsilon$.  Then $I(A';B'|C') \le
I(A;B|C) + 2H_2(\epsilon) + \epsilon\log\min\{|\sA|,|\sB|\}$.
\end{lem}

Note that since $p_{Q'|X',Y'}=p_{Q|X,Y}$, we have $\Delta(XYQ,X'Y'Q') =
\Delta(XY,X'Y') = \epsilon$.  Then we invoke \lemmaref{statdiff-mutualinfo}
thrice (with $(ABC,A'B'C')$ standing for $(YQX,Y'Q'X')$, $(XQY,X'Q'Y')$ and
$(XYQ,X'Y'Q')$, respectively). This combined with the fact that
$\min\{|\sY|,|\sQ|\}$,
$\min\{|\sX|,|\sQ|\}$,
$\min\{|\sX|,|\sY|\}$,
are all upperbounded by $\max\{|\sX|,|\sY|\}$,
we obtain the requisite bounds.

\end{IEEEproof}

\subsection{A Few Examples}
\label{sec:examples}
Obtaining closed form expressions for the region \Rtens XY can be difficult. However, for our
applications it often
suffices to identify parts of the boundary of \Rtens XY. We give a couple of
examples below. A more detailed example appears in \sectionref{example}.

\begin{eg} \label{eg:connected} \figureref{connected} shows the joint
p.m.f. of a pair of dependent random variables $X,Y$. 
\begin{figure}[htb]
\centering
\resizebox{\linewidth}{!}{\input{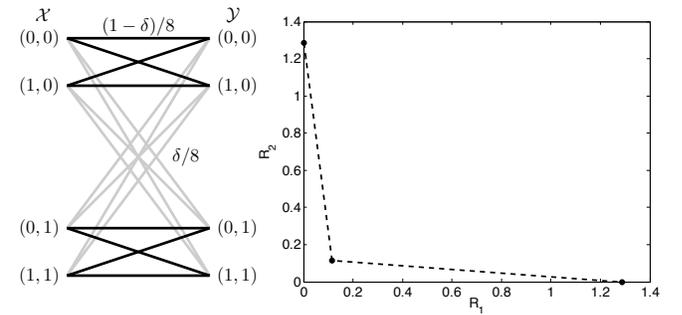}}
\caption{$X,Y$ are dependent random variables whose joint
p.m.f is shown on the left. The solid black lines each carry a probability mass of
$\frac{1-\delta}{8}$ and the lighter ones $\frac{\delta}{8}$. In the plot,
all points $(R_1,R_2)$ on the dotted lines are such that
$(R_1,R_2,0)\in\Rtens XY$.}
\label{fig:connected}
\end{figure}

When $\delta=0$, they
have the simple dependency structure of $X=(X',Q), Y=(Y',Q)$ where
$X',Y',Q$ are independent. This is the perfectly resolvable case. Thus, the
set of rate pairs $(R_1,R_2)$ such that $(R_1,R_2,0)\in\Rtens XY$ is the
entire positive quadrant. For small values of $\delta$ we intuitively
expect the random variables to be ``close'' to this case. A measure such as
the common information of G\'{a}cs and K\"{o}rner fails to bring this out
(common information is discontinuous in $\delta$ jumping from $H(Q)=1$ at
$\delta=0$ to 0 for $\delta>0$). However, the intuition is borne out by our
trade-off regions. For instance, for $\delta=0.05$, \figureref{connected}
shows that the set of rate pairs $(R_1,R_2)$ such that $(R_1,R_2,0) \in
\Rtens XY$ is nearly all of the positive quadrant.
\end{eg}

\begin{eg}{\em A binary example.} \label{eg:zsource} \figureref{zsource}
shows the joint p.m.f. of a pair of dependent binary random variables
$U,V$.  In the plot in \figureref{zsource} we show the intersection of
\Rtens UV with the plane $z=0$. The computation is along the lines of \sectionref{example}.
\begin{figure}[htb]
\centering
\resizebox{\linewidth}{!}{\input{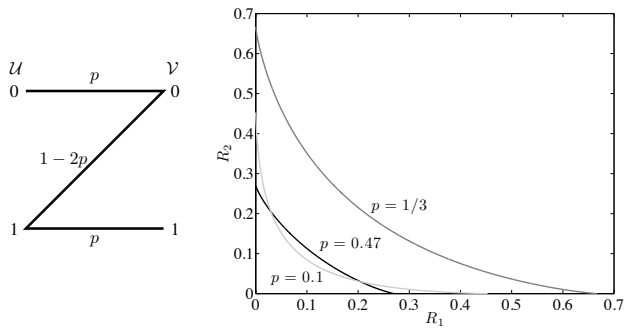}}
\caption{$U,V$ are binary random variables with joint p.m.f.
$p(0,0)=p(1,1)=p$, $p(1,0)=1-2p$, and $p(0,1)=0$.
The plot shows the boundary of the set of all rate pairs $(R_1,R_2)$ such that
$(R_1,R_2,0)\in\Rtens UV$ for three values of $p$. When $p$ approaches 0 or $0.5$,
this boundary approaches the axes, indicating that the random variables are
closer to being perfectly resolvable.}
\label{fig:zsource}
\end{figure}
\end{eg}

\section{Assisted Common Information}
\label{sec:ACI}

\begin{figure}[htb]
\centering
\scalebox{0.6}{\includegraphics{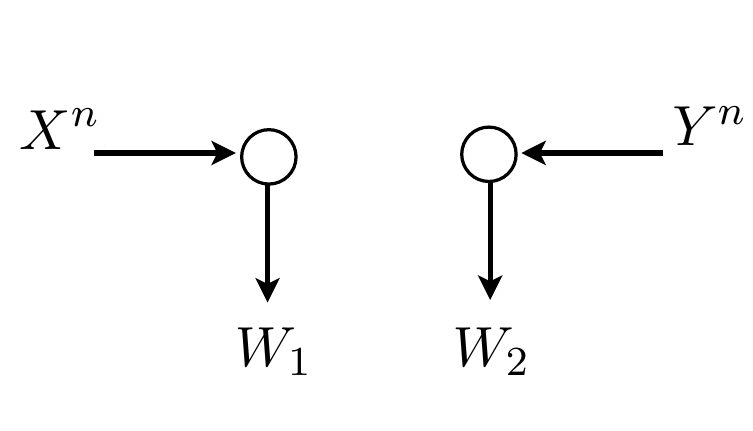}}\\(a)\\\vspace{0.1cm}
\scalebox{0.6}{\includegraphics{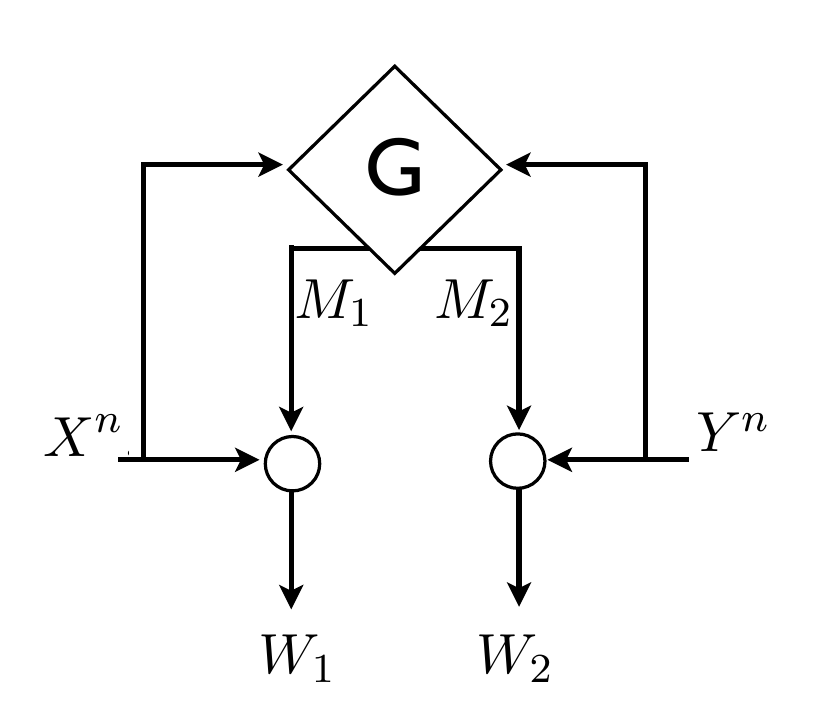}}\\(b)
\caption{(a) Setup for G\'{a}cs-K\"{o}rner common information.
The observers generate $W_1$ and $W_2$ which are required to agree with
high probability. (b) Assisted common information system. 
A genie assists the observers by sending separate messages to
them over rate-limited noiseless links. When the genie is absent the setup
reduces to the one for G\'{a}cs-K\"{o}rner common information.
}
\label{fig:common}
\end{figure}

Recall that
when $X=(X',Q)$ and $Y=(Y',Q)$ where $X',Y',Q$ are independent, then a
natural measure of ``common information'' of $X$ and $Y$ is $H(Q)$. 
In this case, an observer of $X$ and an observer of $Y$ may independently produce the
common part $Q$; and conditioned on $Q$, there is no ``residual
information'' that correlates $X$ and $Y$ i.e., $I(X;Y|Q)=0$. 
The definition $\Com\GKW(X;Y)$ of G\'{a}cs and
K\"{o}rner~\cite{GacsKo73} generalizes this to arbitrary $X,Y$
(\figureref{common}(a)): the two observers now see
$X^n=(X_1,\ldots,X_n)$ and $Y^n=(Y_1,\ldots,Y_n)$, resp., where $(X_i,Y_i)$
pairs are independent drawings of $(X,Y)$. They are required to produce
random variables $W_1=f_1(X^n)$ and $W_2=f_2(Y^n)$, resp., which agree
(with high probability). The largest entropy rate (i.e., entropy normalized
by $n$) of such a ``common'' random variable was proposed as the {\em
common information} of $X$ and $Y$. We will refer to this as the \GKW
common information of $(X,Y)$ and denote it by $\Com\GKW(X;Y)$. However,
in the same paper~\cite{GacsKo73}, G\'{a}cs and K\"{o}rner showed (a result
later strengthened by Witsenhausen~\cite{Witsenhausen75}) that this rate is
still just the largest $H(Q)$ for $Q$ which can be obtained (with
probability 1) as a deterministic function of $X$ alone as well as a
deterministic function of $Y$ alone.
\begin{align*}
\Com\GKW(X;Y) = \max_{\substack{\pQXY:\\H(Q|X)=H(Q|Y)=0}} H(Q).\\
\end{align*}
It is easy to see that the above maximum is achieved by the random variable
$Q$ defined over the set of connected components of the characteristic
bipartite graph of $(X,Y)$, such that $p_{Q|XY}(q|x,y)=1$ if and only if
the edge $(x,y)$ belongs to the connected component $q$.  Note that this
captures only an explicit form of common information in a single instance
of $(X,Y)$.

One limitation of the common information defined by G\'{a}cs and K\"{o}rner is
that it ignores information which is {\em almost} common.%
\footnote{Other approaches which do not necessarily suffer from this
drawback have been suggested,
notably~\cite{Wyner75,AhlswedeKo74,Yamamoto94}. As we show, our
generalization is also intimately connected with~\cite{Wyner75}.}
In particular, if there is only a single connected component in the
characteristic bipartite graph then the common information between them is
zero, even if it is the case that by removing a set of edges that account
for a small probability mass, the graph can be disconnected into a large
number of components each with a significant probability mass.
Our approach in this section could be viewed as a strict generalization of
G\'{a}cs and K\"{o}rner, which uncovers such extra layers of
``almost common information.'' Technically, we introduce an omniscient genie
who has access to both the observations $X^n$ and $Y^n$ and can send separate
messages to the two observers over rate-limited noiseless links.  See
\figureref{common}(b). The objective is for the observers to agree on a
``common'' random variable as before, but now with the genie's assistance. We
call this the {\em assisted common information} system. This leads to a
trade-off region trading-off the rates of the noiseless links and the resulting
common information\footnote{We use the term common information primarily to
maintain continuity with~\cite{GacsKo73}.} (or the resulting residual
mutual information). We characterize these trade-off regions in terms of the region of tension
of the two random variables, and show that, in
general, they exhibit non-trivial behavior, but reduce to the trivial behaviour
discussed above when the rates of the noiseless links are zero.

As before, two observers receive $X^n=(X_1,\dotsc,X_n)$  and
$Y^n=(Y_1,\dotsc,Y_n)$ respectively, and need to output strings $W_1$ and
$W_2$ respectively, that must match each other with high probability. But
here, an omniscient Genie $G$ computes $M_1=f_1\upn(X^n,Y^n)$ and
$M_2=f_2\upn(X^n,Y^n)$ as deterministic functions of $(X^n,Y^n)$ and sends
these to the two observers as shown in \figureref{common}(b). The observers
are allowed to compute their outputs also making use of the respective
messages they receive from the genie, as $W_1=g_1\upn(X^n,M_1)$ and
$W_2=g_2\upn(Y^n,M_2)$, where $g_1\upn$ and $g_2\upn$ are deterministic
functions. Here again, the goal is to study how large the entropy of $W_1$
(and equivalently $W_2$) can be, but controlling for the number of bits
used to transmit $M_1$ and $M_2$.

For a pair of random variables $(X,Y)$ and positive integers $N_1,N_2,n$,
an $(N_1,N_2,n)$ {\em assisted common information} (\aci) {\em code}
is defined as a quadruple $(f_1\upn, f_2\upn, g_1\upn, g_2\upn)$, where
\begin{align*}
f_k\upn&:\sX^n \times \sY^n \rightarrow \{1,\ldots,N_k\},\quad k=1,2\\
g_1\upn&:\sX^n\times\{1,\ldots,N_1\}\rightarrow\bbZ, \text{ and}\\
g_2\upn&:\sY^n\times\{1,\ldots,N_2\}\rightarrow\bbZ 
\end{align*} are deterministic
functions. A sequence of $(N_1(n),N_2(n),n)$ \aci codes
$(f_1\upn,f_2\upn,g_1\upn,g_2\upn)_{n=1,2,\ldots}$ is called a
{\em valid $(R_1,R_2)$ {\rm \aci} strategy} for $(X,Y)$, if for every
$\epsilon > 0$, for sufficiently large $n$,
% there exists an integer $n$ such that
\begin{align}
\frac{1}{n}\log N_k(n) &\leq R_k + \epsilon, \quad k=1,2\label{eq:rates}\\ 
\Pr[ g_1\upn(X^n,f_1\upn(X^n,Y^n)) &\neq g_2\upn(Y^n,f_2\upn(X^n,Y^n))]\notag\\
&\qquad\qquad\qquad\leq \epsilon.
\label{eq:prob-of-error}
\end{align}

We say that a rate pair $(R_1,R_2)$ {\em enables common information rate}
$R_\CI \ge 0$ for $(X,Y)$, if there exists a valid $(R_1,R_2)$ \aci strategy
${(f_1\upn,f_2\upn,g_1\upn,g_2\upn)}_n$ for $(X,Y)$ such that for every
$\epsilon > 0$, for sufficiently large~$n$,
% as in \eqref{eq:prob-of-error},
%
\begin{align} 
\frac{1}{n}H(g_1\upn(X^n,f_1\upn(X^n,Y^n))) \geq R_\CI - \epsilon.
\label{eq:CIrate} 
\end{align}
Similarly, we say that a rate pair $(R_1,R_2)$ {\em enables residual
information rate} $R_\RI$ for $(X,Y)$, if there exists a valid $(R_1,R_2)$
\aci strategy $(f_1\upn,f_2\upn,g_1\upn,g_2\upn)_n$ for $(X,Y)$ such that
for every $\epsilon > 0$, for sufficiently large~$n$,
% as in \eqref{eq:prob-of-error},
%
\begin{align} 
\frac{1}{n} I(X^n;Y^n|g_1\upn(X^n,f_1\upn(X^n,Y^n))) &\leq R_\RI + \epsilon.
\label{eq:RIrate} 
\end{align}
Note that if $(R_1,R_2)$ enables residual information rate $R_{\RI}$, and
$(R_1',R_2',R_\RI')\ge(R_1,R_2,R_\RI)$, then $(R_1',R_2')$ enables residual
information rate $R_\RI'$ too.

\begin{defn}
The {\em assisted common information region} $\sRsACI(X;Y)$ of a pair of
correlated random variables $(X,Y)$ is the set of all $(R_1,R_2,R_\CI)\in
\Rplus^3$ such that $(R_1,R_2)$ enables common information rate $R_\CI$ for
$(X,Y)$. Similarly the {\em assisted residual information rate region}
$\sRsGKW(X;Y)$ of $(X,Y)$ is the set of all $(R_1,R_2,R_\RI)\in \Rplus^3$
such that $(R_1,R_2)$ enables residual information rate $R_\RI$ for
$(X,Y)$. In other words,
\begin{align*}
\sRsACI(X;Y) &\defineqq
\{(R_1,R_2,R_\CI): (R_1,R_2) \text{ enables }\\
&\text{common information rate } R_\CI \text{ for } (X,Y)\},\\
\sRsGKW(X;Y) &\defineqq
\{(R_1,R_2,R_\RI): (R_1,R_2) \text{ enables }\\
&\text{residual information rate } R_\RI \text{ for } (X,Y)\}.
\end{align*}
\end{defn}
We will write $\sRsACI$ and $\sRsGKW$ when the random variables involved
are obvious from the context. It is easy to see from the definition that
$\sRsACI$ and $\sRsGKW$ are closed sets.

Our main results regarding assisted common information system characterize
the assisted residual and common information rate regions of $(X,Y)$, and
relate them to the region of tension of $(X,Y)$.

Recall that $\hsPXY$ is the set of all conditional p.m.f.'s $p_{Q|X,Y}$ such that the
cardinality of alphabet $\sQ$ of $Q$ is such that $|\sQ|\leq |\sX||\sY|+2$.
We have the following characterization of the assisted common and residual information regions:
\begin{thm}\label{thm:GKW-char}
\begin{align*}
&\sRsGKW(X;Y) =
\{(r_1,r_2,r_\RI)\in\Rplus^3:\exists\pQXY\in\hsPXY \text{ s.t.}  \notag\\
&\;\;\;\; r_1\geq I(Y;Q|X), r_2\geq I(X;Q|Y), r_\RI \geq I(X;Y|Q)\}.   \\
&\sRsACI(X;Y) =
\{(r_1,r_2,r_\CI)\in\Rplus^3:\exists\pQXY\in\hsPXY\text{ s.t.}  \notag\\
&\;\;\;\; r_1\geq I(Y;Q|X), r_2\geq I(X;Q|Y), r_\CI \leq I(X,Y;Q)\}. 
\end{align*}
\end{thm}
We prove this theorem in \sectionref{proof}. An immediate consequence is that we have an interpretation of the region of tension $\Rtens XY$ as the assisted residual information region $\sRsGKW(X;Y)$. We may also write it down in terms of the assisted common information region:
\begin{corol}\label{cor:GKW-tension}
For any pair of correlated random variables $(X,Y)$,
\begin{align}
\Rtens XY &= \sRsGKW(X;Y) \label{eq:Rtens-ARI} \\
\Rtens XY &= \ihull{f_{X,Y}(\sRsACI(X;Y))}  \label{eq:Rtens-ACI}
\end{align}
where $f_{X,Y}$ is an affine map defined as
\begin{align*}
f_{X,Y}\left(\left[\begin{array}{c}R_1\\R_2\\R_3\end{array}\right]\right)
\defineqq
 \left[\begin{array}{c}R_1\\ R_2\\I(X;Y) + R_1+R_2 -R_3\end{array} \right].
\end{align*}
\end{corol}
We prove \eqref{eq:Rtens-ACI} in \appendixref{ACI}.

\subsection{Behavior at $R_1=R_2=0$ and Connection to
G\'{a}cs-K\"{o}rner~\cite{GacsKo73}}

As discussed above, G\'{a}cs and K\"{o}rner defined the common information,
$\Com\GKW(X;Y)$ using the system in \figureref{common}(a), where there is
no genie. Formally, an $n$-\gk map-pair $(g_1\upn,g_2\upn)$ is a pair of
maps $g_1\upn:\sX^n\rightarrow \bbZ$ and $g_2\upn:\sY^n\rightarrow \bbZ$.
We will say that $R_\CI$ is an {\em achievable common information rate} for
$(X,Y)$ if there is a sequence of \gk map-pairs
$(g_1\upn,g_2\upn)_{n=1,2,\ldots}$ such that for every $\epsilon>0$, for
large enough $n$,
\begin{align*}
\Pr[g_1\upn(X^n) \neq  g_2\upn(Y^n)] &\leq \epsilon,\text{ and}\\
\frac{1}{n} H\left(g_1\upn(X^n)\right)&\geq R_\CI -\epsilon.
\end{align*}
\gk common information $\Com\GKW(X;Y)$ is the supremum of all achievable
common infomation rates for $(X,Y)$.
As mentioned earlier, G\'{a}cs and K\"{o}rner~\cite{GacsKo73} showed that
$\Com\GKW(X;Y)$ is simply $H(Q)$ where $Q$ corresponds to the connected component
in the characteristic bipartite graph of $(X,Y)$.

It is clear from the definition that $(0,0,\Com\GKW(X;Y))\in\sRsACI(X;Y)$.
However, it is not clear whether $\Com\GKW(X;Y)$ is the
largest value of $R_\CI$ such that $(0,0,R_\CI)\in\sRsACI(X;Y)$;
i.e.,  if we define
$\RACIintz XY$ as the axis intercept of the boundary
of $\sRsACI(X;Y)$ along the $R_\CI$ axis as follows 
\begin{align*}
\RACIintz XY &\defineqq \max  \{R_\CI: (0,0,R_\CI)\in \sRsACI(X;Y)\},
\end{align*}
then it is not immediately clear whether $\Com\GKW(X;Y)=\RACIintz XY$.  This is
because the absence of links from the genie is a more restrictive condition
than allowing ``zero-rate'' links from the genie (notice the $\epsilon$ in
\eqref{eq:rates}). So we may ask whether introducing an omniscient genie,
but with {\em zero-rate} links to the observers, changes the conclusion of
G\'{a}cs-K\"{o}rner. In other words, whether $\RACIintz XY$ is larger than
$\Com\GKW(X;Y)$.  The corollary below  (proven in \appendixref{ACI}) answers
this question in the negative.  Also note that the result of
G\'{a}cs-K\"{o}rner can be obtained as a simple consequence of this
corollary.

\begin{corol} \label{cor:GacsKo}
\begin{align}
\Com\GKW(X;Y) &= \RACIintz XY \label{eq:CGKWequivalence}\\
              &= \max_{\substack{\pQXY\in\sPXY:\\H(Q|X)=H(Q|Y)=0}} H(Q)\label{eq:CGKW}%
.
\intertext{Further,}
\Tintz XY &= I(X;Y) - \Com\GKW(X;Y) \label{eq:CGKWfromRIO}
\end{align}
\end{corol}
Thus, at zero rates for the links, assisted common information exhibits the
same trivial behavior as $\Com\GKW$.

\subsection{Proof of \theoremref{GKW-char}} \label{sec:proof}
We first prove the converse (i.e., $\text{L.H.S.}\subseteq \text{R.H.S.}$). Let
$\epsilon>0$, and $n$ and an $(N_1(n),N_2(n),n)$ \aci code
$(f_1\upn,f_2\upn,g_1\upn,g_2\upn)$ be such that
\eqref{eq:rates}-\eqref{eq:CIrate} hold. Let $C_k=f_k\upn(X^n,Y^n)$, for
$k=1,2$, and $W_1=g_1\upn(X^n,C_1)$ and $W_2=g_2\upn(Y^n,C_2)$. Then,
\begin{align*}
R_1 + \epsilon &\geq \frac{1}{n} H(C_1) \geq \frac{1}{n} H(C_1|X^n)
 \geq \frac{1}{n} H(W_1|X^n)\\
 &\geq \frac{1}{n} I(Y^n;W_1|X^n)\\
 &\namedeqq{a} 
   \frac{1}{n} \sum_{i=1}^n H(Y_i|X_i) - H(Y_i|Y^{i-1},X^n,W_1)\\
 &\geq \frac{1}{n} \sum_{i=1}^n H(Y_i|X_i) -
H(Y_i|X_i,W_1,Y^{i-1},X^{i-1})\\
 &= \sum_{i=1}^n \frac{1}{n} I(Y_i;Q_i|X_i),\\*
 &\qquad\qquad\qquad \text{ } Q_i\defineqq(W_1,Y^{i-1},X^{i-1})\\
 &\namedeqq{b} I(Y_J;Q_J|X_J,J),\\*
 &\qquad\qquad\qquad \text{  }p_J(i)\defineqq\frac1n,i\in\{1,\dotsc,n\},\\ 
 &\namedeqq{c} I(Y_J;Q|X_J),\qquad\qquad\text{  }Q\defineqq(Q_J,J),
\end{align*}
where (a) follows from the independence of $(X_i,Y_i)$ pairs across $i$. In
(b), we define $J$ to be a random variable uniformly distributed
over $\{1,\ldots,n\}$ and independent of $(X^n,Y^n)$. And (c) follows from
the independence of $J$ and $(X^n,Y^n)$. Similarly,
\begin{align}
R_2 + \epsilon &\geq \frac{1}{n} H(C_2|Y^n) \geq \frac{1}{n} H(W_2|Y^n)\notag \\
 &= \frac{1}{n} H(W_1,W_2|Y^n) - \frac{1}{n} H(W_1|W_2,Y^n)\notag \\
 &\geq \frac{1}{n} H(W_1|Y^n) - \frac{1}{n} H(W_1|W_2)\notag\\
 &\namedgeq{a} H(W_1|Y^n) - \kappa\epsilon\notag\\
 &\geq\frac{1}{n} I(X^n;W_1|Y^n) - \kappa\epsilon\label{eq:R2ineq}\\
 &\namedgeq{b} I(X_J;Q|Y_J) - \kappa\epsilon,\notag
\end{align}
where (a) (with $\kappa\defineqq1+\log|\sX||\sY|$) follows from Fano's inequality and
the fact that the range of $g_1$ can be restricted without loss of
generality to a set of cardinality $|\sX|^n|\sY|^n$. And (b) can be shown
along the same lines as the chain of inequalities which gave a lower bound
for $R_1$ above. Moreover,
\begin{align*}
\frac{1}{n} I(X^n;Y^n|W_1) &= \frac{1}{n} \sum_{i=1}^n I(X_i;Y^n|W_1,X^{i-1})\\
 &\geq \frac{1}{n} \sum_{i=1}^n I(X_i;Y_i|W_1,X^{i-1},Y^{i-1})\\
 &= I(X_J;Y_J|Q).
\end{align*}
Since $X_J,Y_J$ has the same joint distribution as $X,Y$, the converse
 for assisted residual
information follows. Similarly, the converse for assisted common
information can be shown using
\begin{align*}
&\frac{1}{n}H(W_1)\\
&\qquad\namedeqq{a}\frac{1}{n} I(X^n,Y^n;W_1) \\
&\qquad = \frac{1}{n} \sum_{i=1}^n H(X_i,Y_i) -  H(X_i,Y_i|W_1,X^{i-1},Y^{i-1})\\
  &\qquad = \frac{1}{n} \sum_{i=1}^n I(X_i,Y_i;Q_i)%\\ &\qquad
  = I(X_J,Y_J;Q),
\end{align*}
where (a) follows from the fact that $W_1$ is a deterministic function of
$(X^n,Y^n)$.
The fact that instead of $\pQXY\in\sPXY$ we can consider
$\pQXY\in\hsPXY$ with alphabet \sQ such that $|\sQ|\leq |\sX||\sY|+2$
follows from Fenchel-Eggleston's strengthening of
Carath\'{e}odory's theorem~\cite[pg. 310]{CsiszarKo81}.

To prove achievability (i.e., $\text{L.H.S.} \supseteq \text{R.H.S.}$), 
we will use a result from lossy source coding. See, e.g.,~\cite[Chapter
10]{CoverT06} for a description of the lossy source coding problem.
Consider a source $p_S$, and source and reconstruction alphabets $\sS$ and
$\shS$, respectively. We have the following lemma:
\begin{lem} \label{lem:testchannel}
Given a conditional distribution $p^\ast_{\hS|S}$, there is a
distortion measure $d:\sS\times\shS\rightarrow\Rplus\cup\{\infty\}$, and a
distortion constraint $D$ such that the $p^\ast_{\hS|S}$ is a minimizer for
\[ R(D) = \min_{p_{\hS|S}:\Exp[p_{S\vphantom{\hS}} p_{\hS|S}]{d(S,\hS)}\leq D} I(S;\hS).\] 
Moreover, unless $I(S;\hS)=0$ (in which case any $d$ works), the distortion
measure $d$ is given by
\begin{align}
d(s,\hs) &= -c\log p^\ast_{S|\hS}(s|\hs) + d_0(s), \label{eq:dist-meas}\\
\intertext{where $c>0$ and the function $d_0$ can be chosen arbitrarily, and}
p^\ast_{S|\hS}(s|\hs) &=
\frac{p^\phast_{S\vphantom{\hS}}(s)p^\ast_{\hS|S}(\hs|s)}
   {\sum_{\ts}p^\phast_{S\vphantom{\hS}}(\ts)p^\ast_{\hS|S}(\hs|\ts)}.
\notag
\end{align}
The distortion constraint $D$ is given by
\begin{align*}
D &= \Exp[p^\phast_{S\vphantom{\hS}}p^\ast_{\hS|S}]{d(S,\hS)}.
\end{align*}
%where the expectation is over $p_{S}p^\ast_{\hS|S}$.
\end{lem}
\begin{proof}
See \cite[Problem 3, pg. 147]{CsiszarKo81}; also see
\cite[Lemma~4]{GastparRV03} for a proof.
\end{proof}

For a given $p^\ast_{Q|XY}\in\hsPXY$, we need to argue that
\begin{align*}
(I(Y;Q|X),I(X;Q|Y),I(X,Y;Q)) &\in \sRsACI(X;Y),\\
(I(Y;Q|X),I(X;Q|Y),I(X;Y|Q)) &\in \sRsGKW(X;Y),
\end{align*}
where the conditional mutual information quantities are evaluated using the
joint distribution $p^\phast_{X,Y}p^\ast_{Q|XY}$. Note that these quantities
are continuous in $p^\ast_{Q|XY}$. Moreover, as was mentioned earlier,
it is easy to verify from their definitions that $\sRsACI(X;Y)$ and
$\sRsGKW(X;Y)$ are closed sets. Hence, we may make the following assumption
on $p^\ast_{Q|XY}$ without loss of generality:\\
{\em Assumption:}
$p^\ast_{Q|XY}(q|x,y)>0$ for all $(x,y,q)\in\sX\times\sY\times\sQ$.\\
In \lemmaref{testchannel}, let $p^\phast_{S}$ be $p^\phast_{X,Y}$ and
$p^\ast_{\hS|S}$ be $p^\ast_{Q|XY}$. Let $d:\sX\times\sY\times\sQ
\rightarrow \Rplus\cup\{\infty\}$ denote the distortion measure and
$D^\ast$ the distortion constraint promised by the lemma.
\begin{align}
 D^\ast = \Exp[p^\phast_{X,Y}p^\ast_{Q|XY}]{d(X,Y,Q)}.\label{eq:Dstar} 
\end{align}
Let
\[d_\mathrm{max} = \max_{\substack{(x,y)\in\sX\times\sY:\\p_{X,Y}(x,y)>0}}
\max_{q\in\sQ} \;d(x,y,q).\]
Under the above Assumption, it is clear from \eqref{eq:dist-meas} that
$d_\mathrm{max} < \infty$.

\begin{figure}[htb]
\centering
\includegraphics[width=0.25\textwidth]{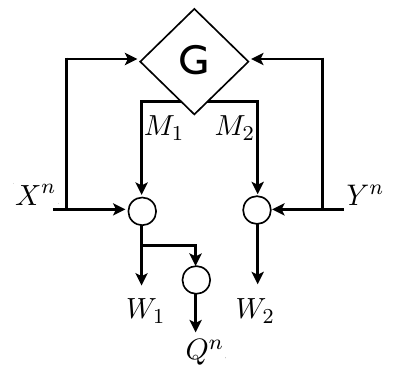}
\caption{Set up in the proof of \theoremref{GKW-char}}
\label{fig:dist-src-coding-problem}
\end{figure}

The rest of the proof proceeds as follows: we will define a distributed
source coding problem (see \figureref{dist-src-coding-problem}) where the
first goal is for the observers to agree on a common random variable as in
the assisted common information setup. However, instead of this common
random variable meeting \eqref{eq:CIrate} or \eqref{eq:RIrate}, we will
require that an output sequence $Q^n$, which is produced as a deterministic
function of the common random variable, must meet a distortion criterion.
The distortion measure and the distortion constraint are those obtained
above using \lemmaref{testchannel}. We will show that these requirements
can be met using a code which operates at $(R_1,R_2) = (I(Y;Q|X),
I(X;Q|Y))$.  We will then argue that this must imply that the common random
variable also meets \eqref{eq:CIrate} and \eqref{eq:RIrate}. 

We make the following definitions (see
\figureref{dist-src-coding-problem}): we define an $(N,N_1,N_2,n)$ {\em
code} as a quintuple $(f_1\upn,f_2\upn,g_1\upn,g_2\upn,h)$, where
\begin{align*}
f_k\upn&:\sX^n \times \sY^n \rightarrow \{1,\ldots,N_k\},\quad k=1,2\\
g_1\upn&:\sX^n\times\{1,\ldots,N_1\}\rightarrow\{1,\ldots,N\},\\
g_2\upn&:\sY^n\times\{1,\ldots,N_2\}\rightarrow\{1,\ldots,N\}, \text{ and}\\
h\upn&: \{1,\ldots, N\} \rightarrow \sQ^n 
\end{align*} are deterministic functions. Note that embedded in this code
is an $(N_1,N_2,n)$ ACI code.
The {\em probability of error} of a code is defined as 
\begin{align}
P\upn_e =\Pr[ &\;g_1\upn(X^n,f_1\upn(X^n,Y^n))\notag\\ 
&\qquad\quad\neq g_2\upn(Y^n,f_2\upn(X^n,Y^n))].\label{eq:Pe}
\end{align}
Let
\[ Q^n = h\upn\left(g_1\upn\left(X^n,f_1\upn(X^n,Y^n)\right)\right).\]
For $D\geq 0$, we will say that $(R_1,R_2,D)$ is {\em achievable} if there
is a sequence of $(N(n),N_1(n),N_2(n),n)$ codes such that for every
$\epsilon > 0$, for sufficiently large $n$,
\begin{align}
\frac{1}{n}\log N_k(n) &\leq R_k+\epsilon, \quad k=1,2
\label{eq:ratescondition}\\
P\upn_e&\leq \epsilon,\label{eq:Pecondition}\\
\intertext{and the following average distortion contraint holds}
\frac{1}{n}\sum_{i=1}^n\Exp{d(X_i,Y_i,Q_i)}&\leq D + \epsilon.\label{eq:dist}
\end{align}
The {\em rate-distortion tradeoff region} $\sR$ is the closure of the set
of all achievable $(R_1,R_2,D)$. 

The following lemma is proved in \appendixref{ACI} using standard
techniques from distributed source coding theory (see, for
instance,~\cite[Chapter~11]{ElGamalKim2012}).
\begin{lem}\label{lem:achievabilitylemma}
\[ (I(Y;Q|X),I(X;Q|Y),D^\ast)\in \sR,\]
where the conditional mututal informations are evaluated using
$p^\phast_{X,Y}p^\ast_{Q|XY}$ and $D^\ast$ is given by \eqref{eq:Dstar}.
\end{lem}

As mentioned above, every code has an ACI code embedded in it. We will show below that if a
code satisfies \eqref{eq:dist} with $D=D^\ast$ of \eqref{eq:Dstar}, then it
must satisfy condition \eqref{eq:CIrate} on common information rate.
More precisely,\\
{\em Claim~1:} If a sequence of $(N(n),N_1(n),N_2(n),n)$ codes satisfy
\eqref{eq:dist} with $D=D^\ast$, then it must hold that for sufficiently
large $n$,
\begin{align*}
\frac{1}{n}H(g\upn_1(X^n,f\upn_1(X^n,Y^n)))&\geq I(X,Y;Q)
-\delta(\epsilon),
\end{align*}
where $\delta(\epsilon)\downarrow 0$ as $\epsilon\downarrow 0$ and the
mutual information expression on the right-hand-side is evaluated using
the joint distribution $p^\phast_{X,Y}p^\ast_{Q|XY}$.
\begin{proof}[Proof of Claim~1]
Suppose \eqref{eq:dist} holds with $D=D^\ast$. Let
$W_1=g\upn_1(X^n,f\upn_1(X^n,Y^n))$. Then,
\begin{align}
H(W_1) &\geq I(W_1;X^nY^n)\notag\\
       &\stackrel{\mathrm{(a)}}{\geq} I(Q^n;X^nY^n)\notag\\
       &= \sum_{i=1}^n I(Q^n;X_iY_i|X^{i-1}Y^{i-1})\notag\\
       &= \sum_{i=1}^n I(Q^nX^{i-1}Y^{i-1};X_iY_i)\notag\\
       &\geq \sum_{i=1}^n I(Q_i;X_iY_i),\label{eq:achievabilitycont}
\end{align}
where (a) is a data processing inequality. Before we proceed further, we
state some simple properties of the rate-distortion function from lossy
source coding:
\begin{align*}
R(D) = \min_{p_{Q|XY}:\Exp{d(X,Y,Q)}\leq D} I(Q;X,Y).
\end{align*}
$R(D)$ is a continuous, convex, and non-increasing function of
$D$.  A proof can be found, for instance, in~\cite{CoverT06}. Let
\begin{align*}
 D_i &= \Exp{d(X_i,Y_i,Q_i)}.\\
\intertext{Then} 
 R(D_i) &\leq I(Q_i;X_iY_i).
\end{align*}
Substituting in \eqref{eq:achievabilitycont},
\begin{align}
H(W_1) &\geq \sum_{i=1}^n R(D_i)\notag\\
  &\stackrel{\mathrm{(a)}}{\geq} n R\left(\frac{1}{n}\sum_{i=1}^n D_i \right)\notag\\
  &\stackrel{\mathrm{(b)}}{\geq} n(R(D^\ast) - \delta(\epsilon)),\label{eq:achievabilitycont2}
\end{align}
where $\delta(\epsilon)\downarrow 0$ as $\epsilon \downarrow 0$. (a) is
Jensen's inequality, and (b) follows from the fact that the code satisfies
\eqref{eq:dist} with $D=D^\ast$ and $R(D)$ is a continuous and
non-increasing function of $D$.

Let us recall that $d$ and $D^\ast$ were provided by \lemmaref{testchannel}
which guarantees that 
\[ R(D^\ast) = I(X,Y;Q), \]
where the mutual information is evaluated using the joint distribution
$p^\phast_{X,Y}p^\ast_{Q|XY}$. Substituting this into 
\eqref{eq:achievabilitycont2} and dividing by $n$, we get Claim~1.
\end{proof}
Further, the conditions \eqref{eq:ratescondition}-\eqref{eq:Pecondition} on
the rates and probability of error of a sequence of codes are identical to
the conditions \eqref{eq:rates}-\eqref{eq:prob-of-error} for a valid ACI
strategy. Hence, we may conclude from \lemmaref{achievabilitylemma} that
\begin{align*}
(I(Y;Q|X),I(X;Q|Y),I(X,Y;Q)) &\in \sRsACI(X;Y).
\end{align*}
To see this, for any $\epsilon'>0$, notice that we may choose a small
enough $\epsilon>0$ such that $\epsilon' \geq \min(\epsilon,
\delta(\epsilon))$.
\lemmaref{achievabilitylemma} promises us an $(N(n),N_1(n),N_2(n),n)$
code such that \eqref{eq:ratescondition}-\eqref{eq:dist}
are met. This implies that \eqref{eq:rates}-\eqref{eq:prob-of-error} are
met with $\epsilon'$. Moreover, Claim~1 implies that \eqref{eq:CIrate} is
also met with $\epsilon'$. This completes the characterization of
$\sRsACI(X;Y)$.

To complete the characterization of $\sRsGKW(X;Y)$, for $\epsilon'>0$, let
$\epsilon>0$ be chosen small enough such that $\epsilon'\geq
(3+\log|\sX||\sY|)\epsilon + \delta(\epsilon)$. Let us consider the $(N(n),
N_1(n), N_2(n), n)$ code promised by \lemmaref{achievabilitylemma} which
satisfies \eqref{eq:ratescondition}-\eqref{eq:dist} with $R_1=I(Y;Q|X)$,
$R_2=I(X;Q|Y)$, and $D=D^\ast$. Let $W_1 = g\upn_1( X^n, f\upn_1(X^n,Y^n))$.
We have the following information theoretic identity (see \eqref{eq:MIeqs4}
on page~\pageref{eq:MIeqs4}): 
\begin{align}
&I(X^n;Y^n|W_1) = I(X^n;Y^n) + I(X^n;W_1|Y^n)\notag\\
 &\qquad\qquad\quad\;\;+ I(Y^n;W_1|X^n) -
I(X^nY^n;W_1). \label{eq:residualachcont}
\end{align}
But,
\begin{align}
I(Y^n;W_1|X^n)&=I(Y^n;g\upn_1(X^n,f\upn_1(X^n,Y^n))|X^n)\notag\\
 &\leq I(Y^n;f\upn_1(X^n;Y^n)|X^n)\notag\\
 &\leq \log N_1(n).\\
\intertext{Using \eqref{eq:Pecondition} and following the same argument
which lead us to \eqref{eq:R2ineq}, we can write} 
I(X^n;W_1|Y^n)&\leq \log N_2(n)+n\kappa\epsilon,\\
\intertext{where $\kappa\defineqq1+\log|\sX||\sY|$. Further, by Claim~1,}
I(X^nY^n;W_1) &= H(W_1)\notag\\ &\geq n(I(X,Y;Q)-\delta(\epsilon)).
\end{align}
Substituting the above three in \eqref{eq:residualachcont} and using
\eqref{eq:ratescondition} with $R_1=I(Y;Q|X)$ and $R_2=I(X;Q|Y)$,
\begin{align}
\frac{1}{n}I(X^n;Y^n|W_1)&\leq I(X;Y) + I(Y;Q|X) + I(X;Q|Y)\notag\\ 
 &\quad - I(X,Y;Q)+ (\kappa+2)\epsilon + \delta(\epsilon)\notag\\
&= I(X;Y|Q) + \epsilon',
\end{align}
where the last equality is again \eqref{eq:MIeqs4}. Hence, we may conclude
that
\[ (I(Y;Q|X),I(X;Q|Y),I(X;Y|Q)) \in \sRsGKW(X;Y).\]
This completes the characterization of $\sRsGKW$.

\section{The Gray-Wyner System and its Relationship to Region of Tension and Assisted Common Information}
\label{sec:GrayWyner}

\subsection{Gray-Wyner system}
\begin{figure}[htb]
\begin{center}
\scalebox{0.6}{\includegraphics{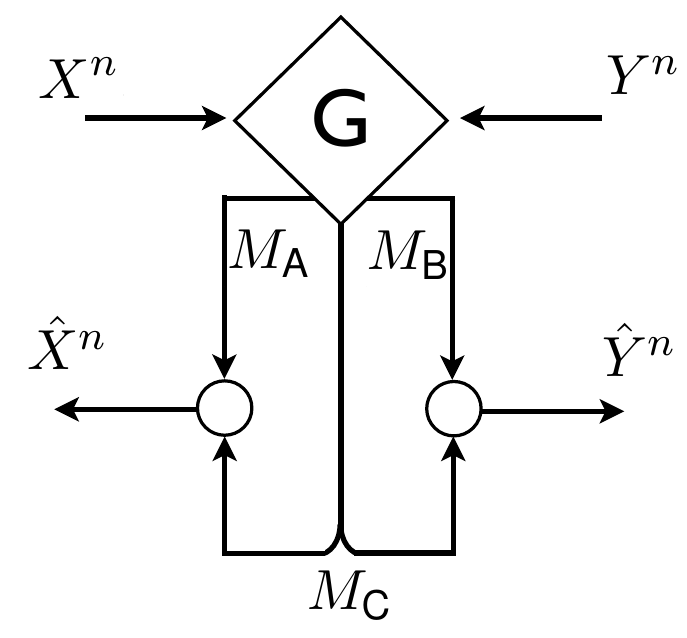}}\\%(b)
\caption{Setup for
Gray-Wyner (GW) system.  }
\label{fig:GW}
\end{center}
\end{figure}

The Gray-Wyner system~\cite{GrayWy74} is shown in \figureref{GW}. It is a source coding problem
where an encoder who observes the pair of correlated sources $X^n,Y^n$ maps it
to three messages: two ``private'' messages $M_\A=f_\A\upn(X^n,Y^n)$,
$M_\B=f_\B\upn(X^n,Y^n)$, and a ``common'' message $M_\C=f_\C\upn(X^n,Y^n)$.
There are two decoders which attempt to recover $X^n$ and $Y^n$ respectively.
The first decoder tries to estimate $X^n$ using the private message $M_\A$ and
the commom message $M_\C$ as $\hX^n=g_{\A\C}\upn(M_\A,M_\C)$, and the second
decoder tries to estimate $Y^n$ from $M_\B,M_\C$ as
$\hY^n=g_{\B\C}\upn(M_\B,M_\C)$. Gray-Wyner problem is to characterize the
rates of the messages so that the decoders estimate losslessly.   

More precisely, for a pair
of random variables $(X,Y)$, an $(N_\A,N_\B,N_\C,n)$ \gw code
$(f_\A\upn,f_\B\upn,f_\C\upn,g_{\A\C}\upn,g_{\B\C}\upn)$, is such that
\begin{align*}
f_\alpha\upn&:\sX^n \times \sY^n \rightarrow
\{1,\ldots,N_\alpha\},\text{ where }
\alpha=\A,\B,\C,\\
g_{\A\C}\upn&:\{1,\ldots,N_\A\}\times\{1,\ldots,N_\C\}\rightarrow\sX^n,
\text{ and }\\
g_{\B\C}\upn&:\{1,\ldots,N_\B\}\times\{1,\ldots,N_\C\}\rightarrow\sY^n
\end{align*}
are deterministic functions. We say that $(R_\A,R_\B,R_\C)$ is {\em
achievable in the Gray-Wyner system} for $(X,Y)$, if there is a sequence of
$(N_\A(n),N_\B(n),N_\C(n),n)$ \gw codes
$(f_\A\upn,f_\B\upn,f_\C\upn,g_{\A,\C}\upn,g_{\B\C}\upn)$ such that for every
$\epsilon > 0$, for large enough $n$
\begin{align*}
\frac{1}{n}\log N_\alpha(n) &\leq R_\alpha + \epsilon,\quad
\alpha=\A,\B,\C,\\
\begin{split}
\Pr[ g_{\A\C}\upn(f_\A\upn(X^n,Y^n), &\,f_\C\upn(X^n,Y^n)) \neq X^n) ] \leq
\epsilon,\\
\Pr[ g_{\B\C}\upn(f_\B\upn(X^n,Y^n), &\,f_\C\upn(X^n,Y^n)) \neq Y^n) ] \leq \epsilon.
\end{split}
\end{align*}
\begin{defn} 
The Gray-Wyner region $\sRsGW(X;Y)$ is the closure of the set of all rate
3-tuples that are achievable in the Gray-Wyner system for $(X,Y)$.
\end{defn}
We write $\sRsGW$ when the random variables are clear from the context.

A simple bound on $\sRsGW(X;Y)$ is given by
$\sRsGW(X;Y) \subseteq \GWlower(X;Y)$, where
\begin{align}
&\GWlower(X;Y) \defineqq \{(R_\A,R_\B,R_\C):
 R_\A+R_\C\geq H(X), \notag\\
&\; R_\B+R_\C\geq H(Y),%\notag\\ &\qquad
R_\A+R_\B+R_\C\geq H(X,Y)\} \label{eq:GWlower}
\end{align}

The Gray-Wyner region was characterized in~\cite{GrayWy74}.
\begin{thm}[\cite{GrayWy74}]\label{thm:GW}
$\sRsGW(X;Y)$ equals 
\begin{align*}
\ihull{ \{(H(X|Q),H(Y|Q),I(X,Y;Q)):  \pQXY\in\hsPXY\}}.
\end{align*}
\end{thm}

Wyner's common information~\cite{Wyner75}, $\Com\Wyner(X;Y)$ of a pair of
random variables $X,Y$ is defined in terms of the Gray-Wyner system. It is
the smallest $R_\C$ such that the outputs of the encoder taken together is
an asymptotically efficient representation of $(X,Y)$, i.e., when
$R_\A+R_\B+R_\C=H(X,Y)$.  Using the above theorem we have
\begin{thm}[\cite{Wyner75}]
\label{thm:Wyner}
\begin{align*}
\Com\Wyner(X;Y) &\defineqq
\inf_{\substack{(R_\A,R_\B,R_\C)\in \sRsGW(X;Y), \\ R_\A+R_\B+R_\C=H(X,Y) }} R_\C \\
       &= \min_{\substack{\pQXY\in\sPXY:\\X-Q-Y}} I(X,Y;Q)
\end{align*}
\end{thm}

It is known that 
G\'acs-K\"orner common information can be obtained from
the Gray-Wyner region~\cite[Problem 4.28, pg. 404]{CsiszarKo81}.
\setcounter{MYtempeqncounter1}{\value{equation}}
\begin{align}
\Com \GKW(X;Y) &= \max_{\substack{R_\A+R_\C=H(X), R_\B+R_\C =H(Y),\\
                                 (R_\A,R_\B,R_\C)\in \sRsGW }} R_\C
\label{eq:oldCGKW}
\end{align}
Alternatively~\cite{KamathAn10},
\begin{align}
\Com \GKW(X;Y) &= \max_{\substack{R \leq I(X;Y),\\ \{R_\C=R\} \cap \GWlower \subseteq \sRsGW}} R \label{eq:AltCGKW}
\end{align}

\subsection{New Connections}\label{sec:GW-Tension}
Analogous to \corollaryref{GKW-tension}, the following theorem (proved in
the appendix) shows that the region of tension of $(X,Y)$ can be expressed
in terms of their Gray-Wyner region. 
\begin{thm} \label{thm:affine}
\begin{align*}
\Rtens XY &= \ihull{g_{X,Y}(\sRsGW(X;Y))},\\
\intertext{where $g_{X,Y}$ is an affine map defined as}
g_{X,Y}\left(\left[\begin{array}{c}R_\A\\R_\B\\R_\C\end{array}\right]\right)
&\defineqq
 \left[\begin{array}{c}
	R_\A+R_\C-H(X)\\
	R_\B+R_\C-H(Y)\\
	R_\A+R_\B+R_\C-H(X,Y)
 \end{array} \right].
\end{align*}
\end{thm}

Thus, the tension region $\Rtens XY$ is the increasing hull of the
Gray-Wyner region $\sRsGW(X;Y)$ under an affine map $g_{X,Y}$. The map, in
fact, computes the gap of $\sRsGW(X;Y)$ to the simple lower bound
$\GWlower(X;Y)$ of \eqref{eq:GWlower}.
The first coordinate of $\sRsGW'$ is the gap between
the (sum) rate at which the first decoder in the Gray-Wyner system receives
data and the minimum possible rate at which it may receive data so that it
can losslessly reproduce $X^n$. The second coordinate has a similar
interpretation with respect to the second decoder. The third coordinate is
the gap between the rate at which the encoder sends data and the minimum
possible rate at which it may transmit to allow both decoders to losslessly
reproduce their respective sources.

Though \theoremref{affine} shows that the region of tension is closely
related to the Gray-Wyner region, it must be noted that the latter does not
possess an essential monotonicity property of the region of tension that
is discussed in \sectionref{crypto}, and is therefore less-suited for the
cryptographic application which motivates this paper.  

The relations \eqref{eq:oldCGKW} and \eqref{eq:AltCGKW} 
fall out of \theoremref{affine} and \corollaryref{GacsKo}.
\begin{corol}\label{cor:CGKWfromThm}
\setcounter{MYtempeqncounter}{\value{equation}}
\setcounter{equation}{\value{MYtempeqncounter1}}
\begin{align}
\Com \GKW(X;Y) &= \max_{\substack{R_\A+R_\C=H(X), R_\B+R_\C =H(Y),\\
                                 (R_\A,R_\B,R_\C)\in \sRsGW }} R_\C \\
\Com \GKW(X;Y) &= \max_{\substack{R \leq I(X;Y),\\ \{R_\C=R\} \cap \GWlower \subseteq \sRsGW}} R 
\end{align}
%restore the current equation number.
\setcounter{equation}{\value{MYtempeqncounter}}
\end{corol}
Another consequence of \theoremref{affine} is an expression for Wyner's
common information $\Com \Wyner(X;Y)$ in terms of $\Rtens XY$ (see
\figureref{Rtens}):
\begin{corol}\label{cor:CGWfromThm}
\begin{align}
\Com \Wyner(X;Y) = I(X;Y) + \min_{(R_1,R_2,0)\in\Rtens XY} R_1 + R_2.
\label{eq:CWynerfromRtens}
\end{align}
\end{corol}

As we have seen already, one of the axes intercepts of $\Rtens XY$, namely
$\Tintz XY$ is closely connected to the \gk common information ($\Com
\GKW(X;Y) = I(X;Y) - \Tintz XY$). The other two axes intercepts also turn
out to be closely connected to certain quantities identified elsewhere in the
context of source coding~\cite{MarcoEf09,KamathAn10}. Before we look at this
connection, let us reinterpret these two axes intercepts using the
fact that $\Rtens XY = \sRsGKW(X;Y)$ (\corollaryref{GKW-tension}). 

In the context of the assisted common information system in
\figureref{common}(b), $\Tintx XY$ (resp., $\Tinty XY$) is the rate at
which the genie must communicate when it has a link to only the user who
receives $X$ (resp.  $Y$) source so that the users can produce a common
random variable conditioned on which the sources are
independent\footnote{Though the definition allows for zero-rate
communication to the other user and a zero-rate (but non-zero) residual
conditional mutual information, it can be shown from the expression for
these rates in \eqref{eq:corner1}-\eqref{eq:corner2} that there is a scheme
which achieves exact conditional independence and requires no communication
to the other user. The proof is similar to that of \corollaryref{GacsKo}.}.
We have already seen in \theoremref{intercepts} that
\begin{align}
\Tintx XY &= \min_{\substack{\pQXY\in\sPXY:\\I(X;Q|Y)=I(X;Y|Q)=0}} I(Y;Q|X), \label{eq:corner1}\\
\Tinty XY &= \min_{\substack{\pQXY\in\sPXY:\\I(Y;Q|X)=I(X;Y|Q)=0}} I(X;Q|Y). \label{eq:corner2}
\end{align}

We will show below that this pair is closely related to a
pair of quantities identified in the context of lossless coding with
side-information~\cite{MarcoEf09} and the Gray-Wyner
system~\cite{KamathAn10}. Let (following the notation of~\cite{KamathAn10}) 
\begin{align*}
&G(Y\rightarrow X)=\\&\;\;\min\{R_\C:(H(X|Y),H(Y)-R_\C,R_\C) \in\sRsGW(X;Y)\},\\
&G(X\rightarrow Y)=\\&\;\;\min\{R_\C:(H(X)-R_\C,H(Y|X),R_\C) \in \sRsGW(X;Y)\}.
\end{align*}
It has been shown~\cite{MarcoEf09,KamathAn10} that $G(Y\rightarrow X)$ is
the smallest rate at which side-information $Y$ may be coded and sent
to a decoder which is interested in recovering $X$ with asymptotically
vanishing probability of error if the decoder receives $X$ coded and sent
at a rate of only $H(X|Y)$ (which is the minimum possible rate which will
allow such recovery). Further, \cite{KamathAn10} arrives at the maximum of
$G(Y\rightarrow X)$ and $G(X\rightarrow Y)$ as a dual to the alternative
definition of $\Com \GKW$ in \eqref{eq:AltCGKW} from the Gray-Wyner system.

We prove the following relationship between the two pairs of quantities in
the appendix.
\begin{corol}\label{cor:cornerconnection}
\begin{align}
G(Y\rightarrow X) &= I(X;Y) + \Tintx XY,\label{eq:cornerresult1}\\
G(X\rightarrow Y) &= I(X;Y) + \Tinty XY.\label{eq:cornerresult2}
\end{align}
Further,
\begin{align}
&\min\{R:R\geq I(X;Y),\notag\\ &\qquad\;\;\;\;\;(R_\C=R)\cap\GWlower(X;Y) \subseteq \sRsGW(X;Y)\}\notag\\
&\qquad\qquad=\max(G(Y\rightarrow X),G(X\rightarrow Y))\label{eq:kamathdual}\\
&\qquad\qquad=I(X;Y)+\max(\Tintx XY, \Tinty XY)\label{eq:kamathdualresult}.
\end{align}
\end{corol}

\section{Upperbounds on the Rate of Two-Party Secure Sampling
Protocols}
\label{sec:crypto}

We will now apply the concept of tension to derive
upperbounds on the rate of two-party secure sampling protocols.
A two-party protocol $\Pi$ is specified by a pair of (possibly randomized)
functions \pialice and \pibob, that are used by each
party to operate on its current state $W$ to produce a message $m$ (that is
sent to the other party) and a new state $W'$ for itself. The initial state
of the parties may consist of correlated random variables $(X,Y)$, with
Alice's state being $X$ and Bob's state being $Y$; such a pair is called a 
{\em set up} for the protocol.
The protocol proceeds by the parties taking turns to apply their respective
functions to their state, and sending the resulting message to the other
party; this message is added to the state of the other party.
$\pi_{\mathrm{Alice}}$ and $\pi_{\mathrm{Bob}}$ also specify when the
protocol terminates and produces output (instead of producing the next
message in the protocol).
A protocol is considered {\em valid} only if both parties
terminate in a finite number of rounds (with probability 1).
The {\em view} of a party in an execution of the protocol is a random
variable which is defined as the sequence of its states so far in the
protocol execution.
For a valid protocol $\Pi=(\pialice,\pibob)$, we shall denote the final
views of the two parties as $(\Pialice(X;Y),\Pibob(X;Y))$. Also, we shall
denote the outputs as $(\Pialiceout(X;Y),\Pibobout(X;Y))$.
(Later, when it is clear, we abbreviate these as
$(\Pialice,\Pibob)$ and $(\Pialiceout,\Pibobout)$
respectively.)

Now we define (perfectly) secure sampling. (Extension to statistically
secure sampling, which allows a vanishing error, is treated
in \sectionref{statistical}.)

\begin{defn}
\label{def:secure-sampling}
We say that a pair of correlated random variables $(U,V)$
can be (perfectly) {\em securely sampled} using a pair
of correlated random variables $(X,Y)$ as set up if
there exists a valid protocol $\Pi=(\pialice,\pibob)$
such that
\begin{align}
&(\Pialiceout(X;Y),\Pibobout(X;Y))  \sim p_{U,V},
			\label{eq:correctness} \\
&\Pialice(X;Y) \mchain \Pialiceout(X;Y) \mchain \Pibobout(X;Y), 
			\label{eq:Bsecure} \\
&\Pialiceout(X;Y) \mchain \Pibobout(X;Y) \mchain \Pibob(X;Y) 
			\label{eq:Asecure}
\end{align}
are Markov chains.
In this case we say \samples\Pi{(X,Y)}{(U,V)}.
\end{defn}

The three conditions above correspond to correctness (when neither party is
corrupt), security for Bob when Alice is corrupt, and security for Alice
when Bob is corrupt. The correctness condition in \eqref{eq:correctness} is obvious:
 the outputs $(\Pialiceout(X;Y),\Pibobout(X;Y))$ must be identically
distributed as $(U,V)$. 
The condition in \eqref{eq:Bsecure} says that even if Alice is ``curious''
(or ``passively corrupt'') and retains her view in the entire protocol,
it should give her no more
information about Bob's output than 
just her own output at the end of the protocol
provides. \eqref{eq:Asecure} gives the symmetric condition for when
Bob is curious.

Before proceeding, we remark that a basic question regarding secure sampling
is to characterize the random variables $(U,V)$ which can be securely
sampled {\em without any set up}. Note that if $(U,V)$ is perfectly
resolvable -- i.e., there is a random variable $Q$ such that
$H(Q|U)=H(Q|V)=0$ and $I(U;V|Q)=0$ -- then there is a simple protocol for
securely sampling $(U,V)$: Alice samples $Q$ and sends it to Bob, and then
Alice and Bob privately sample $U$ and $V$ respectively, conditioned on $Q$.
In fact, these are the only random variables which have secure
sampling protocols, even if we allow a relaxed notion of security
(\definitionref{stat-secure-sampling}).
\begin{prop}
\label{pro:trivial}
$(U,V)$ has a statistically secure sampling protocol without any set up,
if and only if $(U,V)$ is perfectly resolvable.
\end{prop}
This result, for the case of perfect security follows for instance, from \cite{WolfWu08};
for the case of statistical security, it follows as a special case
of the bound presented below in \corollaryref{tension-stat-rate-bound}.

\subsection{Towards Measuring Cryptographic Content}
As metioned in \sectionref{tension}, in \cite{WolfWu08} three information
theoretic quantities were introduced, which we identified as the three axes
intercepts of \Rtens XY.  As shown in \cite{WolfWu08}, these quantities are
``monotones'' that  can only decrease in a protocol, and if the protocol
securely realizes a pair of correlated random variables $(U,V)$ using a set
up $(X,Y)$, then each of these quantities should be at least as large for
$(X,Y)$ as for $(U,V)$. Thus such a monotone can be thought of as a quantitative
measure of cryptographic content in the sense that 
$(U,V)$ with a higher cryptographic content cannot be
generated from a set up $(X,Y)$ with a lower cryptographic content.
In particular, it can be used to bound the ``rate'' $n_1/n_2$ so that
$n_1$ independent copies of $(U,V)$ can be generated from $n_2$ independent
copies of $(X,Y)$ (as defined later, in \definitionref{sampling-rate}).

While the quantities in \cite{WolfWu08} do capture several interesting cryptographic
properties, they paint a very incomplete picture. For instance, two pairs of
correlated random variables $(X,Y)$ and $(X',Y')$ may have vastly different
values for these quantities, even if they are statistically close to each
other, and hence have similar ``cryptographic content.'' %
In \cite{WinklerWu10}, (among other things) this was addressed to some
extent by extending some of the bounds in \cite{WolfWu08} to statistical
security. However, these results still considered separate monotones, with
no apparent relationship with each other.

\vspace{0.04cm}
Instead, we shall consider a single three dimensional region $\Rtens XY$ and show that the
region as a whole satisfies a monotonicity property: the region can only expand (grow towards
the origin) when $(X,Y)$ evolve as the views of the two parties in a
protocol (or outputs ``securely derived'' from the views in a protocol).
Hence if the protocol securely realizes a pair of correlated random
variables $(U,V)$ using a set up $(X,Y)$, then \Rtens XY should be contained
within \Rtens UV.  As we shall see, since the region \Rtens XY has a non-trivial shape
(see for instance, \exampleref{zsource}), \Rtens XY can yield much better bounds on the
rate than just considering the axis intercepts; in particular \Rtens XY can
differentiate between pairs of correlated random variables that have the
same axis intercepts.  Further \Rtens XY is continuous as a function of
$p_{X,Y}$, and as such one can derive rate bounds  that are applicable to
statistical security as well as perfect security. Our bounds 
improve over those in \cite{WolfWu08,WinklerWu10}, and as illustrated in
\sectionref{example}, can give interesting tight bounds which evaded the
previous techniques.

\subsection{Monotone Regions for 2-Party Secure Protocols}
\label{sec:monotone}

\begin{defn}
\label{def:monotone}
We will call a function \Mfunc that maps a pair of random variables $X$ and
$Y$, to an upward closed subset%
\footnote{A subset \Mfunc of $\Real^d$ is called upward closed if $\pt \in
\Mfunc$ and $\pt' \ge \pt$ (i.e., each co-ordinate of $\pt'$ is no less than that
of \pt) implies that $a'\in \Mfunc$.}
of $\Rplus^d$ (points in
the $d$-dimensional real space with non-negative co-ordinates)  a {\em monotone region}
if it satisfies the following properties:
\begin{enumerate}
\item ({\em Local computation cannot shrink it.})
For all jointly distributed random variables $(X,Y,Z)$ with $X \mchain Y
\mchain Z$, we have
$ \M{XY}{Z} \supseteq \M{Y}{Z}$ and $\M{X}{YZ} \supseteq \M{X}{Y}$.
\item ({\em Communication cannot shrink it.})
For all jointly distributed random variables $(X,Y)$ and functions $f$ (over the support of
$X$ or $Y$), we have
$ \M{X}{Yf(X)} \supseteq \M{X}{Y}$ and $\M{Xf(Y)}{Y} \supseteq \M{X}{Y}$.
\item  ({\em Securely derived outputs do not have smaller regions.})
For all jointly distributed random variables $(X,U,V,Y)$ with $X \mchain U \mchain V$ and
$U\mchain V \mchain Y$, we have
$ \M{U}{V} \supseteq \M{XU}{YV}$.
\item ({\em Regions of independent pairs add up.})
For independent pairs of jointly distributed random variables $(X_1,Y_1)$ and $(X_2,Y_2)$, we
have
$ \M{X_1X_2}{Y_1Y_2} = \M{X_1}{Y_1} + \M{X_2}{Y_2}, $
where the $+$ sign denotes {\em Minkowski sum}. In other words,
$\M{X_1X_2}{Y_1Y_2} = \{ \pt_1+\pt_2 \;|\; \pt_1 \in \M{X_1}{Y_1} 
 \text{ and } \pt_2\in\M{X_2}{Y_2} \}$.
(Here addition denotes coordinate-wise addition.) 
\end{enumerate}
\end{defn}
Note that since \M{X_1}{Y_1} and \M{X_2}{Y_2} have non-negative co-ordinates
and are upward closed, $\M{X_1}{Y_1} + \M{X_2}{Y_2}$ is smaller than both of
them.  This is consistent with the intuition that more cryptographic content
(as would be the case with having more independent copies of the random
variables) corresponds to a smaller region.

Our definition of a monotone region  strictly generalizes that suggested by
\cite{WolfWu08}. The monotone in \cite{WolfWu08}, which is
a single real number $m$, can be interpreted as a one-dimensional region
$[m,\infty)$ to fit our definition. (Note that a decrease in the value of
$m$ corresponds to the region $[m,\infty)$ enlarging.)

\begin{thm} \label{thm:monotone-rate-bound}
If $n_1$ independent copies of a pair of correlated random variables $(U,V)$ can be securely realized
using $n_2$ independent copies of a pair of correlated random variables $(X,Y)$ as set up, then
for any monotone region \Mfunc,
$n_2 \M XY \subseteq n_1 \M UV$.
(Here multiplication by an integer $n$
refers to $n$-times repeated Minkowski sum.)
\end{thm}
\begin{IEEEproof}
Consider some protocol $\Pi$ such that
$\samples\Pi{(X^{n_2},Y^{n_2})}{(U^{n_1},V^{n_1})}$. Let $t$ be
the maximum number of messages in the protocol.
For $i=0,\dotsc,t$, let $(X_i,Y_i)$ denote the views of
the parties after the $i^\text{th}$ message. Then $(X_0,Y_0)=(X^{n_2},Y^{n_2})$
and $(X_t,Y_t)=(\Pialice,\Pibob)$.
By Condition~(1) and Condition~(2)
of \definitionref{monotone}, $\M {X_{i+1}}{Y_{i+1}} \supseteq \M {X_i}{Y_i}$
(note that we do allow the local computation defined by \pialice and \pibob
to be randomized, but the randomness used is independent of the other party's
view). By \eqref{eq:correctness}-\eqref{eq:Asecure} as
applied to
$\samples\Pi{(X^{n_2},Y^{n_2})}{(U^{n_1},V^{n_1})}$,
and Condition~(3), 
$\M {U^{n_1}}{V^{n_1}} = \M \Pialiceout\Pibobout \supseteq \M {X_t}{Y_t}$.
Thus, $\M {U^{n_1}}{V^{n_1}}  \supseteq \M {X^{n_2}}{Y^{n_2}}$.
Finally, by Condition~(4) we obtain the claimed inclusion.
\end{IEEEproof}

\subsection{Using Tension to Bound Rate of Secure Sampling}
\label{sec:Rtens-monotone}

\theoremref{monotone-rate-bound} gives us a means to use 
an appropriate monotone region
to bound the {\em rate} of securely sampling instances of a pair $(U,V)$ from a
set up $(X,Y)$. We define this rate as follows (where $(X^n,Y^n)$
denotes $n$ independent copies of $(X,Y)$).
\begin{defn}
\label{def:sampling-rate}
For pairs of correlated random variables $(U,V)$ and $(X,Y)$ (i.e., p.m.f.s $p_{UV}$ and $p_{XY}$),
the {\em rate of securely sampling $(U,V)$ from $(X,Y)$} is defined as%
\footnote{Here we let $\frac{n_1}{n_2}=0$ when $n_1=n_2=0$.}
\[ \sup \{ \frac{n_1}{n_2} : \exists \Pi,n_1,n_2 \text{ s.t. } \samples\Pi{(X^{n_2},Y^{n_2})}{(U^{n_1},V^{n_1})} \}. \]
\end{defn}

Note that in \theoremref{monotone-rate-bound}, $n$-times repeated Minkowski sum of \Mfunc is 
\begin{align*}
n \Mfunc = \{ \pt_1+\cdots+\pt_n \;|\; \pt_1,\dotsc,\pt_n \in \Mfunc \}.
\end{align*}
In general, the shape of the $n$-times Minkowski sum of a region changes with $n$ and
would make it difficult to work with.
But if \Mfunc is convex, then this multiplication operation gives the same
region as the following definition of
multiplication by a real number $r> 0$:
\begin{align}
r \cdot \Mfunc = \{ r \pt \;|\; \pt \in \Mfunc \} &&\text{(for convex $\Mfunc$)}.
\label{eq:convex-mult}
\end{align}
This gives us a convenient way to bound the rate, if we use a
{\em convex monotone region}.
The following is an immediate corollary of 
\theoremref{monotone-rate-bound} (and the fact that for convex regions
$\Mfunc_1$ and $\Mfunc_2$,
$n_2 \Mfunc_2 \subseteq n_1 \Mfunc_1$
iff 
$ \Mfunc_2 \subseteq \frac{n_1}{n_2} \Mfunc_1$).
\begin{corol}
\label{cor:convex-monotone-rate-bound}
For any convex monotone region \Mfunc,
if the rate of securely sampling $(U,V)$ from $(X,Y)$ is $r>0$, then
$\M XY \subseteq r \cdot \M UV$. (Here, multiplication of a region
by a real number is as in \eqref{eq:convex-mult}.)
\end{corol}

The importance of the above corollary is that the region of tension 
provides us with a ``good'' convex monotone region, which can be used to obtain state-of-the-art
bounds on the rate.
\begin{thm}\label{thm:Rtens-monotone}
\Kfunc is a (3-dimensional) monotone region (as in \definitionref{monotone}).
\end{thm}
In fact, we shall show a more general result in
\theoremref{Rtens-robust-monotone}, which implies the above theorem.
Combined with the fact that \Kfunc is convex
(\theoremref{Rtens-convex}), \theoremref{Rtens-monotone}
and \corollaryref{convex-monotone-rate-bound} yield the following result
(which will also be generalized in \corollaryref{tension-stat-rate-bound}).
\begin{corol}
\label{cor:tension-rate-bound}
If the rate of securely sampling $(U,V)$ from $(X,Y)$ is $r>0$, then
$\Rtens XY \subseteq r \cdot \Rtens UV$. 
\end{corol}
Note that this gives an {\em upperbound} on $r$, because, as $r$ increases
from 0, the region $r\cdot\Rtens XY$ shrinks away from the origin.

In general, we can obtain tighter bounds this way than yielded by the three
monotones considered in \cite{WolfWu08} (namely, the axis intercepts of this
monotone region), because the region of tension can ``bulge'' towards the
origin.  In other words, the intercepts, and in particular the common
information of G\'{a}cs and K\"{o}rner, do not by themselves capture subtle
characteristics of correlation that are reflected in {\em the shape of the
monotone region}. Below, we give a concrete example where the region
of tension does give us a tighter bound than the monotones of
\cite{WolfWu08}.

\begin{eg} \label{eg:crypto-example} Consider the question of securely
realizing $n_1$ independent pairs of random variables distributed according
to $(U,V)$ in \exampleref{zsource} from $n_2$ independent pairs of $(X,Y)$
in \exampleref{connected}.  While the monotones in \cite{WolfWu08} will give
an upperbound of $1.930$ on the rate $n_1/n_2$, we show that $n_1/n_2 \le 0.551$. (For
this we use the intersection of \Rtens UV with the plane $z=0$
(\figureref{zsource}) and one point in the region \Rtens XY (marked in
\figureref{connected}); then by \corollaryref{tension-rate-bound},
$0.1143 \ge 0.2075 \cdot r$. Note that we do not claim this is the tightest bound we
can obtain from \corollaryref{tension-rate-bound}: we do not check if
$\Rtens XY \subseteq r\cdot\Rtens UV$  for this value of $r$, since we do not compute the entire 
boundary of the two three-dimensional regions.)
\end{eg}

\subsection{Statistical Security}
\label{sec:statistical}

Recall that the security conditions 
(\eqref{eq:correctness}--\eqref{eq:Asecure})
for 
a protocol $\Pi$ sampling $(U,V)$ from a
set up $(X,Y)$
relate $\Pialiceout(X;Y),\Pibobout(X;Y),\Pialice(X;Y),\Pibob(X;Y)$
with $U,V$ and with each other. These conditions
are for {\em perfect security}. A more realistic
notion of security allows a small error in all these
three conditions. Such a notion is referred to as
{\em statistical security}. Below, we present a standard ``simulation-based''
definition of statistical security. (Below, we will abbreviate
$\Pialiceout(X;Y),\Pialice(X;Y)$ etc.~by
$\Pialiceout,\Pialice$ etc., for the sake of readability.)

\begin{defn}
\label{def:stat-secure-sampling}
For $\epsilon\geq0$, a protocol $\Pi$ is said to $\epsilon$-{\em securely sample}
a pair of correlated random variables $(U,V)$
using a pair
of correlated random variables $(X,Y)$ as set up if
there exists a valid protocol $\Pi=(\pialice,\pibob)$
and random variables (``simulated views'') \simPialice
and \simPibob, over the alphabets  of
\Pialice and \Pibob respectively, distributed according to $p_{\simPialice|U,V}$ and
$p_{\simPibob|U,V}$ such that
\begin{align}
\simPialice \mchain U \mchain V &\qquad\text{ and }\qquad U \mchain V \mchain \simPibob 
\label{eq:stat-simulation}\\
\Delta\big(~ \big(U,V\big), &~ \left(\Pialiceout,\Pibobout\right) ~\big) \le \epsilon
\label{eq:stat-correctness}\\
\Delta\big(~\left(\simPialice,V\right), &~ \left(\Pialice,\Pibobout\right) ~\big) \le \epsilon 
\label{eq:stat-Bsecure}\\
\Delta\big(~\left(U,\simPibob\right), &~ \left(\Pialiceout,\Pibob\right) ~\big) \le \epsilon
\label{eq:stat-Asecure}
\end{align}
Here $\Delta(\cdot,\cdot)$ stands for the total variation distance.
In this case we say \statsamples\epsilon\Pi{(X,Y)}{(U,V)}.
\end{defn}

\paragraph*{Remark}  \statsamples0\Pi{(X,Y)}{(U,V)} if and only if
\samples\Pi{(X,Y)}{(U,V)} (\definitionref{secure-sampling}). In particular, it can be shown that if
\statsamples0\Pi{(X,Y)}{(U,V)},
then \eqref{eq:Bsecure} and \eqref{eq:Asecure} hold
(see for instance, \lemmaref{simulation-to-smalldependency}).
In the other direction, if \samples\Pi{(X,Y)}{(U,V)}, then one can
take $p_{\simPialice|U,V}=p_{\Pialice|\Pialiceout,\Pibobout}$ and
$p_{\simPibob|U,V}=p_{\Pibob|\Pialiceout,\Pibobout}$.

\begin{defn}
\label{def:stat-secure-sampling-rate}
We say  $(U,V)$ can be {\em statistically securely sampled} using a pair
of correlated random variables $(X,Y)$ as set up if, for any $\epsilon >0$,
there is a valid protocol $\Pi$ and positive integers $n_1,n_2$ such that 
$\statsamples\epsilon\Pi{(X^{n_2},Y^{n_2})}{(U^{n_1},V^{n_1})}$.
Then, the {\em rate of statistically securely sampling $(U,V)$ from $(X,Y)$} is defined as
\[ \lim_{\epsilon\downarrow 0} \sup
\left\{ \frac{n_1}{n_2} : \exists \Pi,n_1,n_2 \text{ s.t. } \statsamples\epsilon\Pi{(X^{n_2},Y^{n_2})}{(U^{n_1},V^{n_1})} \right\}.
\]
\end{defn}

\paragraph*{Remark}The typical definition of security in cryptography
literature requires the protocol $\Pi$ to be {\em uniform} (i.e., the
protocol for all values of $\epsilon$ can be implemented by a single Turing
Machine that takes $\epsilon$ as input) and also ``efficient'' (i.e., the
Turing Machine implementing the protocol runs in time (say) polynomial in
$\log\nicefrac1\epsilon$). Since we shall be proving negative results, using
the weaker security definitions without these restrictions only strengthens
our results.

\paragraph*{Robust Monotone Regions} We generalize the definition of a monotone
region (\definitionref{monotone}) by strengthening item~(3) in the definition
to the following conditions, to obtain the definition of a ``robust monotone
region.''

\begin{defn}
\label{def:robust-monotone}
We will call a function \Mfunc that maps a pair of random variables $X$ and
$Y$, to an upward closed subset of $\Rplus^d$  a {\em robust monotone
region} if it is a monotone region (as in \definitionref{monotone}),
and the following hold:
\begin{enumerate}
\item[3$'$)]  ({\em Statistically securely derived outputs do not have a much smaller region.})
There exists a constant $c\ge 0$ such that,
for any jointly distributed random variables $(X,U,V,Y)$ 
and $\phi\ge0$, if
$I(X;V|U)\le\phi$ and $I(U;Y|V)\le\phi$,
then 
\[\M{U}{V} \supseteq \M{XU}{YV} + c\phi.\]

\item[3$''$)]  
({\em Continuity, Convexity and Closure.}) 
There exists a bounded, continuous function
$\del: [0,1] \rightarrow \Rplus$
with $\del(0)=0$, such that 
for any two pairs of correlated random variables $(X,Y)$ and $(X',Y')$,
both over alphabet $\sX \times \sY$,
and $\epsilon\in[0,1]$, if $\Delta(XY,X'Y')= \epsilon$, then
$\M {X}{Y} \subseteq \M {X'}{Y'} -\del(\epsilon)\cdot\log|\sX||\sY|$.
Also,
\M XY is convex and closed.
\end{enumerate}
\end{defn}
Note that condition (3) in \definitionref{monotone} is a restriction of
condition (3$'$) to the case $\phi=0$. 

In \appendixref{crypto} we prove the following generalization
of \corollaryref{convex-monotone-rate-bound}.
\begin{thm}
\label{thm:robust-monotone-stat-rate-bound}
For any robust monotone region \Mfunc,
if the rate of statistically securely sampling $(U,V)$ from $(X,Y)$ is $r>0$, then
$\M XY \subseteq r \cdot \M UV$.
\end{thm}

Also, we can generalize \theoremref{Rtens-monotone} as follows.
\begin{thm}\label{thm:Rtens-robust-monotone}
\Kfunc is a (3-dimensional) robust monotone region (as in \definitionref{robust-monotone}).
\end{thm}
\begin{IEEEproof}
We verify the four properties of a robust monotone region (see
\definitionref{monotone} and \definitionref{robust-monotone}).
\begin{asparaenum}
\item {\em Local computation cannot shrink it:} For all random variables with
 $X - Y - Z$, we need to show that $\Rtens X {YZ} \supseteq \Rtens X Y$ and $\Rtens {XY} Z
\supseteq \Rtens X Y$.

The first inclusion follows from the fact that for the joint p.m.f.
$p_{XYZQ}=p_{XY}p_{Z|Y}p_{Q|XY}$, we have
\begin{align*}
I(X;YZ|Q) &= I(X;Y|Q)\\
I(Q;YZ|X) &= I(Q;Y|X)\\
I(X;Q|YZ) &= I(X;Q|Y).
\end{align*}
\item {\em Communication cannot shrink it:} For all random variables $(X,Y)$
and functions $f$ over the support of $X$ (resp, $Y$), we have to show that $\Rtens X
{(Y,f(X))} \supseteq \Rtens X Y$ (resp, $\Rtens {(X,f(Y))} Y \supseteq \Rtens X Y$).

The first set inclusion follows from the following facts for the joint
p.m.f $p_{XYZQ}=p_{XY}p_{Z|Y}p_{Q|XY}$:
\begin{align*}
I(X;Y,f(X)|Q,f(X))&=I(X;Y|Q,f(X))\\&\leq I(X;Y|Q)\\
I(X;Q,f(X)|Y,f(X))&=I(X;Q|Y,f(X))\\&\leq I(X;Q|Y)\\
I(Y;Q,f(X)|X)&=I(Y;Q|X).
\end{align*} 
The second inclusion follows analogously.
\item[3$'$)] {\em Statistically securely derived outputs do not have a much smaller region:}
We let $c=1$.
Suppose 
$I(X;V|U)\le\phi$ and $I(U;Y|V)\le\phi$.
We shall show that $\Rtens UV \supseteq \Rtens {XU}{VY} + \phi$.
For this, it is enough to show that, for any $\pQXUVY \in \sPXUVY$, 
$\tens UVQ \le \tens{XU}{VY}Q + \phi$
(where the comparison is coordinate-wise and the
addition applies to each coordinate).
This is easy to see for the last coordinate since
$I(U;V|Q) \le I(XU;VY|Q) \le I(XU;VY|Q)+\phi$.
For the second coordinate, note that
\begin{align*}
I(XU;Q|VY) &\ge I(U;Q|VY) \\ 
&= I(U;QY|V)- I(U;Y|V) \\
&\ge I(U;Q|V)- I(U;Y|V).
\end{align*}
Since 
$I(U;Y|V)\le\phi$, we have
$I(U;Q|V) \le I(XU;Q|VY) + \phi$.
Similarly,
$I(V;Q|U) \le I(VY;Q|XU) + \phi$.

\item[3$''$)] 
Continuity follows from \theoremref{Rtens-continuity}, with 
$\del(\epsilon)=2H_2(\epsilon)+\epsilon$
(so that $\delta(\epsilon)$ in \theoremref{Rtens-continuity} is upper-bounded
by $\del(\epsilon)\log|\sX||\sY|$). 
Convexity and closure follow from \theoremref{Rtens-convex} and \theoremref{Rtens-closed}
respectively.

\item[4)] {\em Regions of independent pairs add up:}
If $(X_1,Y_1)$ is independent of $(X_2,Y_2)$, we have to show that $\Rtens
{(X_1X_2)} {(Y_1Y_2)} = \Rtens {X_1} {Y_1} + \Rtens {X_2} {Y_2}$. This follows easily from the following
facts:

For the joint p.m.f. $p_{X_1 Y_1}p_{X_2 Y_2}p_{Q_1|X_1 Y_1}p_{Q_2|X_2 Y_2}$, we have 
\begin{align*}
I(X_1 X_2;Y_1 Y_2|Q_1 Q_2)&=I(X_1;Y_1|Q_1) + I(X_2 Y_2|Q_2),\\
I(X_1 X_2;Q_1 Q_2|Y_1 Y_2)&=I(X_1;Q_1|Y_1) + I(X_2;Q_2|Y_2),\\
I(Y_1 Y_2;Q_1 Q_2|X_1 X_2)&=I(Y_1;Q_1|X_1) + I(Y_2;Q_2|X_2).
\end{align*}
From this, it follows that \[\Rtens
{(X_1X_2)} {(Y_1Y_2)} \supseteq \Rtens {X_1} {Y_1} + \Rtens {X_2} {Y_2}.\]

To show  inclusion in the other direction, consider a joint p.m.f. $p_{X_1 Y_1}p_{X_2 Y_2}p_{Q|X_1 Y_1 X_2 Y_2}$. Let $Q_1=Q$ and $Q_2=Q X_1 Y_1$. Then we have
\begin{align*}
I(X_1 X_2;Y_1 Y_2|Q) =&~ I(X_1; Y_1 | Q) +  I(X_2; Y_1 | Q X_1) \\
& + I(X_1; Y_2 | Q Y_1)  + I(X_2; Y_2 | Q_2) \\
\geq&~ I(X_1;Y_1|Q_1) + I(X_2;Y_2|Q_2).
\end{align*}
Also,
\begin{align*}
I(X_1 X_2;Q|Y_1 Y_2) 
=&~ H(X_1 |Y_1) + H(X_2|Y_2) \\
&- H(X_1|Q Y_1 Y_2) - H(X_2 | Q_2 Y_2 ) \\
\geq&~ H(X_1|Y_1) + H(X_2|Y_2) \\
& - H(X_1|Q Y_1 ) - H(X_2 | Q_2 Y_2 ) \\
=&~ I(X_1;Q_1|Y_1) + I(X_2;Q_2|Y_2).
\end{align*}
Similarly,
\begin{align*}
I(Y_1 Y_2;Q|X_1 X_2)
\geq I(Y_1;Q_1|X_1) + I(Y_2;Q_2|X_2).
\end{align*}
\end{asparaenum}
\end{IEEEproof}

\theoremref{robust-monotone-stat-rate-bound} and
\theoremref{Rtens-robust-monotone} together yield a generalization of \corollaryref{tension-rate-bound}.
\begin{corol}
\label{cor:tension-stat-rate-bound}
If the rate of statistically securely sampling $(U,V)$ from $(X,Y)$ is $r>0$, then
$\Rtens XY \subseteq r \cdot \Rtens UV$. 
\end{corol}

\subsection{Bounding the Rate of Bit-OT from String-OT} \label{sec:example}

\exampleref{crypto-example} was contrived to highlight the shortcomings
of prior work.  We now give another example where the upperbound from our
result strictly improves on prior work, but is further interesting for two
reasons: firstly, the new example is based on natural correlated random
variables that are widely studied (namely, variants of oblivious transfer), and
secondly, the new upperbound we can prove actually matches an easy lowerbound
and is therefore tight.

\paragraph*{Bit-Oblivious Transfer and String-Oblivious Transfer} Oblivious
Transfer, or OT~ \cite{Rabin81,Wiesner83}, is a  pair of correlated random
variables with great cryptographic significance. There are several variants
of OT that have been considered in the literature. In particular, ``bit-OT''
corresponds to the following correlated pair of random variables:
$A=(S_1,S_2)$ and $B=(C,S_C))$ where $S_1,S_2$ are two i.i.d.~uniformly
random bits and the ``choice bit'' $C$ is independent of $(S_1,S_2)$ and
takes a uniformly random values in $\{1,2\}$. Informally, in bit-OT, one of
the two bits that Alice gets is transferred to Bob, but Alice is oblivious
to which one was chosen to be transferred.

It is well-known that all non-trivial correlated random variables (i.e.,
those for which the tension region excludes the origin), including the
different forms of OT, are all ``qualitatively equivalent,'' in the sense
that one can be securely sampled using another as set up \cite{Kilian00}.
However, the rate at which this can be done has not been studied well. That
these rates are non-zero follows from a recent result in
\cite{IshaiKuOsSa09}. We are interested in upperbounding this rate (and
indeed, when possible, calculating it exactly).

Consider the rate of sampling bit-OT from a generalization of bit-OT called
``string-OT'' where Alice receives two $L$-bit strings $S_1,S_2$ instead of
two bits (and one of those strings is obliviously transmitted to Bob).  It
is not hard to see that the rate of sampling bit-OT from string-OT is 1,
intuitively because a single instance of string-OT provides only one bit $C$
that is hidden from Alice. (In terms of the monotones, the axis intercept
$\Tintx AB = (1,0,0)$ for string-OT, independent of the length of the
strings.) But what if we consider two string-OTs together, one in each
direction? In this case, there are $L$ bits with Bob that are hidden from
Alice, and vice versa. We ask if we can sample OT from this set up at a rate
larger than 1 (in particular, linear in $L$).

Formally, we consider the set up $(X,Y)$ and target random variables $(U,V)$
as defined below.

Let $S_{A,1},S_{A,2},S_{B,1},S_{B,2} \in \{0,1\}^L$ and $C_A,C_B
\in \{1,2\}$ be six independent random variables all of which are uniformly
distributed over their alphabets.  Consider a pair of random variables
$X,Y$ defined as $X=(C_A,S_{A,1},S_{A,2},S_{B,C_A})$ and
$Y=(C_B,S_{B,1},S_{B,2},S_{A,C_B})$. 
(Note that $(S_{A,1},S_{A,2},C_A)$ and $(S_{B,1},S_{B,2},C_B)$ correspond
to the two instances of $L$-bit string-OT, one in each direction.)
Let
$U,V$ be a pair of random variables whose joint distribution is the same as
that of $X,Y$, but with $L=1$. In other words, $U,V$ are a pair of
independent bit-OT's in opposite directions. (This is in fact, equivalent
to two independent copies of bit-OT's in the same direction, as can be seen
from the symmetry of the characteristic bipartite graph of bit-OT, which
is simply an 8-cycle \cite{WolfWu06}.)

It is easy to see that $\KK XY$ intersects the coordinate axes at
$(1+L,0,0)$, $(0,1+L,0)$, and $(0,0,2L)$. From, these we can immediately
obtain the upperbound of~\cite{WolfWu08} on the rate, namely
$(1+L)/2$. Notice that this is dependent on $L$ and would suggest that
(several) long string-OT pairs can be turned into several (more) bit-OT
pairs. However, as we show below, the rate is just 1,
i.e., the best one can do is to turn each pair of string-OT's into a pair
of bit-OT's. (This also means that the rate at which bit-OT's can be
obtained per pair of string-OT's is 2, since a pair of bit-OT's in opposite
directions is identical to a pair of bit-OT's in the same direction.)

To see this we need to consider a point on \Rtens XY other than the three
axis intercepts. By setting $Q=(C_A,C_B,S_{A,C_B},S_{B,C_A})$ we get $\tens
XYQ =(1,1,0)$; that is, \Rtens XY contains a point $(1,1,0)$ independent of
$L$. This already bounds the rate of sampling $(U,V)$ from $(X,Y)$ as set
up, by some constant. To show that this constant is 1, we shall show that
$(1,1,0)$ occurs {\em on the boundary of \Rtens UV}.  Then it follows from
\corollaryref{tension-stat-rate-bound} that the rate of (statistically)
secure sampling is upperbounded by 1.

To show that $(1,1,0)$ occurs on the boundary of \Rtens UV, we show that
$\inf\{R_1+R_2: (R_1,R_2,0)\in\KK UV\} = 2$. Since
$\KK UV$ is a monotone region (\theoremref{Rtens-monotone}), by property (4) of
\definitionref{monotone}, the regions of independent pairs add up, Hence, we
need only characterize the $\inf\{R_1+R_2:(R_1,R_2,0)\in \KK AB\}$, where
$(A,B)$ is a single pair of independent bit-OT's: 
$A=(S_1,S_2)\in\{0,1\}^2$ uniformly distributed over its alphabet and
$B=(C,S_C)$, where $C\in\{1,2\}$ is independent of $A$ and uniformly distributed
over its alphabet.
\begin{align*}
&\inf\{R_1+R_2:(R_1,R_2,0)\in\KK AB\}\\
&\quad= \inf_{p_{Q|AB}\in\sPXY:I(A;B|Q)=0} I(B;Q|A) + I(A;Q|B)\\ 
&\quad= H(A|B)+H(B|A)\\
&\qquad\quad - \sup_{p_{Q|AB}\in\sP:I(A;B|Q)=0} H(A|QB) + H(B|QA).
\end{align*}
We show below that the $\sup$ term is 1. Since $H(A|B)+H(B|A)=2$, this will
allow us to conclude that the smallest sum-rate $R_1+R_2$ such that
$(R_1,R_2,0)\in\KK AB$ is 1. Invoking the lemma above, the
corresponding smallest sum-rate for $U,V$ is then 2 as required.

\begin{figure*}
\centering
\subfloat[]{\scalebox{0.4}{\includegraphics{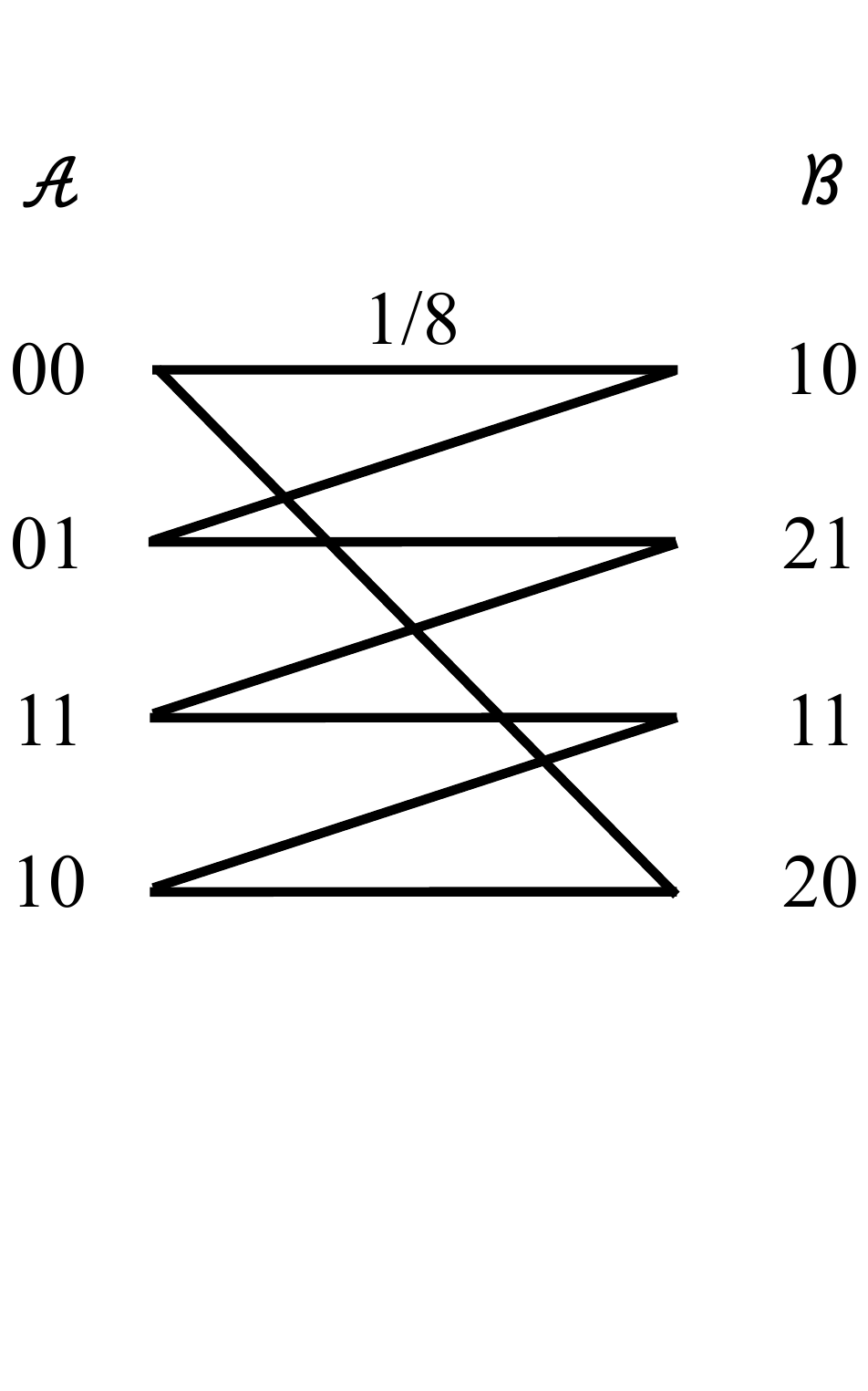}}}\qquad\qquad%
\subfloat[]{\scalebox{0.35}{\includegraphics{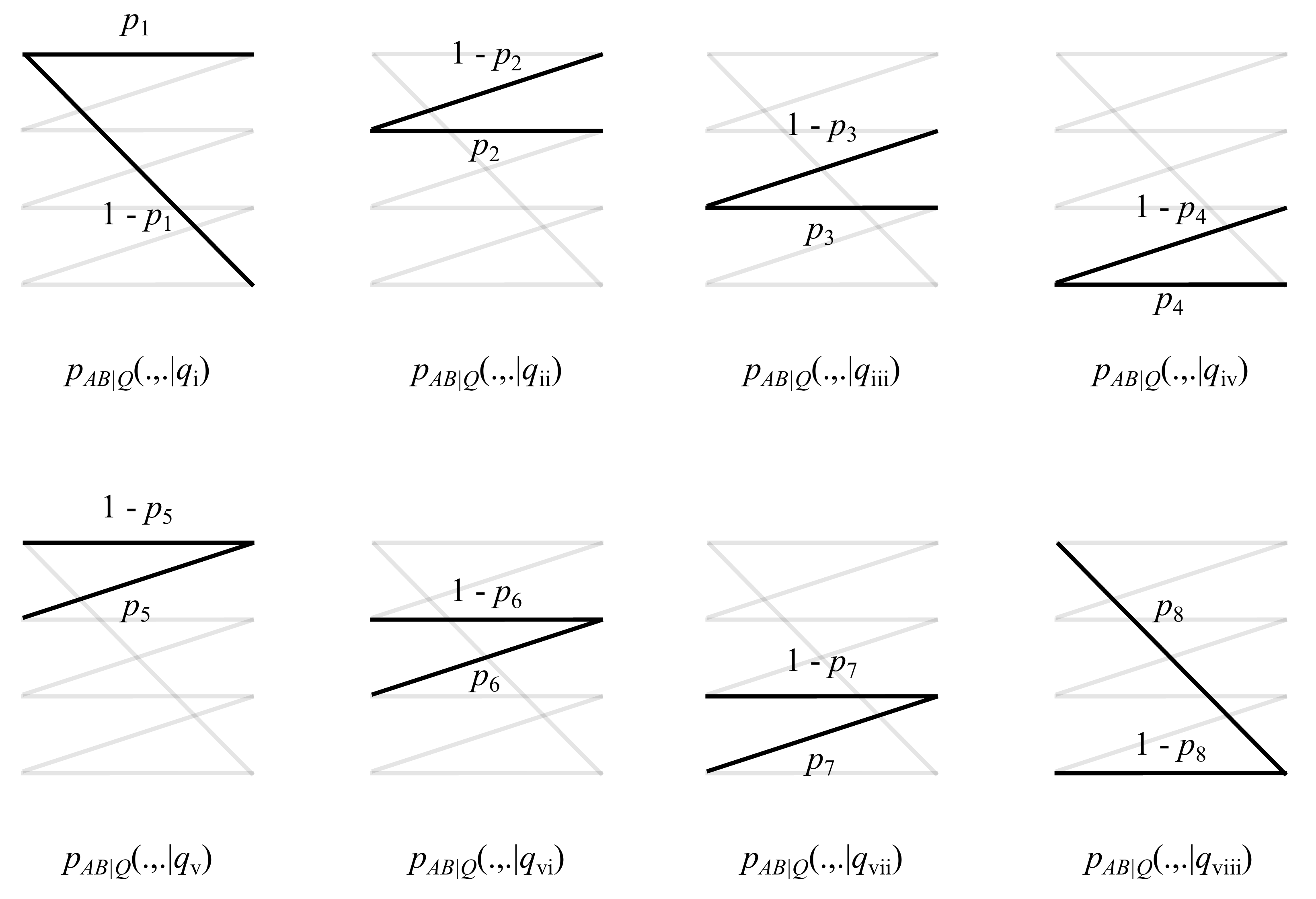}}%
            \label{fig:example-cond-pmf}}
\caption{(a) Joint p.m.f. of $A,B$. Each solid line represents a probablity
mass of 1/8. (b) Eight possible classes that $p_{AB|Q}(.,.|q)$ may belong
to for a $p_{Q|AB}$ which satisfies $I(A;B|Q)=0$.}
\label{fig:example}
\end{figure*}

To show that the $\sup$ term is 1, notice that the only valid choices of
$p_{Q|AB}$ are such that $I(A;B|Q)=0$. This means that the resulting
$p_{AB|Q}(\cdot,\cdot|q)$ must belong to one of eight possible classes shown in
\figureref{example-cond-pmf} (for any $q$ with non-zero probability
$p_Q(q)$; we may assume that all $q$'s have non-zero probability without
loss of generality). Recall that there is a cardinality bound on $Q$; let
us denote the alphabet of $Q$ by $\{q_1,q_2,\ldots,q_N\}$, where $N$ is the
cardinality bound. 

We will first show that there is no loss of generality in assuming that no
more than one of the $q_i$'s is such that its $p_{AB|Q}(.,.|q_i)$ belongs
to the same class (and hence we may take $N=8$). Suppose, $q_1$ and $q_2$
belong to the same class, say class~1, with parameters $p_1$ and $p_2$
respectively.  Then, if we denote the binary entropy function by $H_2(.)$,
we have
\begin{align*}
&H(A|QB) + H(B|QA)\\
&= \sum_{k=1}^N p_Q(q_k) \left( H(A|BQ=q_k) + H(B|AQ=q_k) \right)\\
&= p_Q(q_1) H_2(p_1) + p_Q(q_2) H_2(p_2)\\
   &\qquad + \sum_{k=3}^N p_Q(q_k) \left(H(A|BQ=q_k) + H(B|AQ=q_k)\right)\\
&\leq \left(p_Q(q_1)+p_Q(q_2)\right)
   H_2\left(\frac{p_Q(q_1)p_1 + p_Q(q_2)p_2}{p_Q(q_1)+p_Q(q_2)}\right)\\
  &\qquad + \sum_{k=3}^N p_Q(q_k) \left( H(A|BQ=q_k)+H(B|AQ=q_k)\right),
\end{align*}
where the inequality (Jensen's) follows from the concavity of the binary
entropy function. Thus, we can define a $Q'$ of alphabet size $N-1$ where
letters $q_1,q_2$ are replaced by $q_0$ such that
$p_{Q'}(q_0)=p_{Q}(q_1)+p_{Q}(q_2)$, and $p_{AB|Q'=q_0}$ is in class~1
with parameter $\frac{p_Q(q_1)p_1 + p_Q(q_2)p_2}{p_Q(q_1)+p_Q(q_2)}$, while
maintaining for $i=3,\ldots,N$, $p_{Q'}(q_i)=p_Q(q_i)$ and
$p_{AB|Q'}(a,b|q_i)=p_{AB|Q}(a,b|q_i)$. (It is easy to verify (a) that
this gives a valid joint p.m.f. for $p_{ABQ'}$, (b) that the induced
$p_{AB}$ is the same as the original, and (c) that the induced $p_{Q'|AB}$
satisfies the condition $I(A;B|Q')=0$.) Then, the above inequality states
that
\[ H(A|QB) + H(B|QA) \leq H(AQ'B)+H(B|Q'A)\]
proving our claim.

Thus, without loss of generality, we may assume that $N=8$ and
$p_{AB|Q}(\cdot,\cdot|q_i)$ belongs to class~$i$. Notice that 
\begin{align*}
p_{Q|AB}(q_1|00,10) + p_{Q|AB}(q_5|00,10) &= 1,
%&& 
\\
p_{Q|AB}(q_2|01,10) + p_{Q|AB}(q_5|01,10) &= 1,
\\
p_{Q|AB}(q_2|01,21) + p_{Q|AB}(q_6|01,21) &= 1,
%&& 
\\
p_{Q|AB}(q_3|11,21) + p_{Q|AB}(q_6|11,21) &= 1,
\\
%\intertext{\ldots}
p_{Q|AB}(q_3,11,11) + p_{Q|AB}(q_7|11,11) &=1,
%&& 
\\
p_{Q|AB}(q_4|10,11) + p_{Q|AB}(q_7|10,11) &= 1,
\\
p_{Q|AB}(q_4|10,20) + p_{Q|AB}(q_8|10,20) &= 1,
%&& 
\\
p_{Q|AB}(q_1|00,20) + p_{Q|AB}(q_8|00,20) &= 1.
\end{align*}
Let us define 
\begin{align*}
\tp_1 &\defineqq p_{Q|AB}(q_1|00,10),&&
\tp_5 \defineqq p_{Q|AB}(q_5|01,10),\\
\tp_2 &\defineqq p_{Q|AB}(q_2|01,21),&&
\tp_6 \defineqq p_{Q|AB}(q_6|11,21),\\
%\intertext{\ldots}
\tp_3 &\defineqq p_{Q|AB}(q_3|11,11),&&
\tp_7 \defineqq p_{Q|AB}(q_7|10,11),\\
\tp_4 &\defineqq p_{Q|AB}(q_4|10,20),&&
\tp_8 \defineqq p_{Q|AB}(q_8|00,20).
\end{align*}

Let us evaluate $H(B|QA)$ in terms of the above parameters. Notice that
$H(B|Q=q_i,A)=0$ for $i=5,\ldots,8$. Hence
\begin{align*}
&H(B|QA)\\
&=\sum_{\begin{subarray}{c}(q,a)\in\{(1,00),(2,01),\\
                           \qquad\;\;\;\,(3,11),(4,10)\}
        \end{subarray}} p_{QA}(q,a)H(B|Q=q,A=a)\\
 &= \frac{\tp_1+(1-\tp_8)}{8}H_2\left(\frac{\tp_1}{\tp_1+(1-\tp_8)}\right)\\
   &\qquad+ 
    \frac{\tp_2+(1-\tp_5)}{8}H_2\left(\frac{\tp_2}{\tp_2+(1-\tp_5)}\right)\\
   &\qquad 
    +\frac{\tp_3+(1-\tp_6)}{8}H_2\left(\frac{\tp_3}{\tp_3+(1-\tp_6)}\right)\\
   &\qquad+ 
    \frac{\tp_4+(1-\tp_7)}{8}H_2\left(\frac{\tp_4}{\tp_4+(1-\tp_7)}\right)\\
 &\leq \frac{4+\sum_{i=1}^4\tp_i - \sum_{j=5}^8 \tp_j}{8},
\end{align*}
where the inequality follows from the fact that binary entropy function is
upperbounded by 1. Similary, we can get
\begin{align*}
H(A|QB) \leq \frac{4+\sum_{j=5}^8\tp_j - \sum_{i=1}^4\tp_i}{8}.
\end{align*}
Combining, we obtain, as desired,
\[ H(B|QA) + H(A|QB) \leq 1.\]

\paragraph*{Remark} Note that we have actually shown that for bit-OT
$(A,B)$, the intersection of \Rtens AB on the plane $z=0$ is the increasing
hull of the line segment between $(1,0,0)$ and $(0,1,0)$. This follows from
what we showed above (i.e., $\inf\{R_1+R_2:(R_1,R_2,0)\in\KK AB\} = 1$) combined
with the fact that $\Tintx AB = (1,0,0)$ and $\Tinty AB = (0,1,0)$, and that
\Rtens AB is convex.

\section{Conclusion}
\label{sec:conclusion}
In this work, we introduced a multi-dimensional measure of correlation
between two random variables, called the region of tension. We show
that the region of tension yields an exact characterization of
the rate-region of a 3-party communication problem, that extends
the 2-party problem considered by G\'acs and K\"orner \cite{GacsKo73}.

Further, relying on a monotonicity property of the region of tension in {\em
secure protocols}, we show that the region of tension can be used to derive
lowerbounds on the rate of securely sampling a pair of correlated random
variables, using samples from another joint distribution as a setup.  While
we use this to obtain tight bounds for secure sampling in many problems, we
leave open the question of whether there are cases where the bounds derived
from the region of tension are loose.  Another open problem is to derive
tight lowerbounds for secure {\em computation}.  We note that while bounds
for secure sampling do yield bounds for secure computation, they tend to be
loose, in general.

As defined here, the region of tension is for two correlated random
variables. We leave it open to devise analogous notions for more than two
parties, with analogous applications. (One such notion, applicable to a
specialized context, was defined in \cite{PrabhakaranPr12}.) Other potential
directions of study include extending the region of tension to the setting
of quantum information, and the possibility of basing the definition of
tension on quantities other than mutual information.

\section*{Acknowledgements}
The example in \sectionref{example} is based on a suggestion by J\"{u}rg
Wullschleger. The first author would like to gratefully acknowledge
discussions with Venkat Anantharam, P\'eter G\'{a}cs, and Young-Han Kim. We
thank Hemanta Maji and Mike Rosulek for discussions at an early stage
in this work.  We also thank Suhas Diggavi and the anonymous referees for carefully reviewing
our drafts and making several insightful comments that have helped us
greatly improve the paper.

%\newpage

\appendices

\begin{figure*}[!t]\label{float:affine}
\normalsize
The following simple information theoretic identities for three jointly
distributed random variables $X,Y,Q$ are used at several places in this paper.
\begin{align}
I(Y;Q|X) &= I(X Y;Q) - I(X;Q) = H(X|Q) + I(X Y;Q) -
H(X),\label{eq:MIeqs1}\\
I(X;Q|Y) &= I(X Y;Q) - I(Y;Q) = H(Y|Q) + I(X Y;Q) - H(Y),\label{eq:MIeqs2}\\
I(X;Y|Q) &= H(X|Q) + H(Y|Q) - H(X Y|Q) 
          = H(X|Q) + H(Y|Q) + I(X Y;Q) - H(X Y),\label{eq:MIeqs3}\\
I(X;Y|Q) &= I(X;Y) + I(Y;Q|X) + I(X;Q|Y) - I(XY;Q). \label{eq:MIeqs4}.
\end{align}
The first three equalities are easy to follow. The last one can be obtained
by subtracting the first two from the third.

\hrulefill
\vspace*{4pt}
\end{figure*}

\section{Details Omitted from \sectionref{tension}}
\label{app:tension}
\begin{lem}[See Problem~3.4.25 in page~402 of~\cite{CsiszarKo81}]
\label{lem:connectedcomp}
Given a pair of random variables $(X,Y)$ and a p.m.f.~\pQXY such that
$I(Y;Q|X)=I(X;Q|Y)=0$, there exists a p.m.f.~$p_{Q'|XY}$   such that
$H(Q'|X)=H(Q'|Y)=0$ and $Q-Q'-XY$.
\end{lem}
\begin{IEEEproof}
Suppose \pQXY is such that $I(Y;Q|X)=I(X;Q|Y)=0$. Then
\begin{align*}
p_{Q|XY}(q|x,y) &= p_{Q|X}(q|x)= p_{Q|Y}(q|y).
\end{align*}
Hence, for all $(x,y)$ such that $p_{XY}(x,y)>0$, we must have
$\forall q$, $p_{Q|X}(q|x)=p_{Q|Y}(q|y)$. This implies that, in the
characteristic bipartite graph (which has vertices in $\sX\cup\sY$ and an edge between $x\in\sX$
and $y\in\sY$ if and only if $p_{XY}(x,y)>0$), for 
each connected component $C\subseteq \sX\cup\sY$, there is a distribution
$p^C_Q$ such that for all $x\in C\cap\sX$ and all $q$,
$p_{Q|X}(q|x)=p^C_Q(q)$; similarly, for all $y\in C\cap\sY$ and all $q$,
$p_{Q|Y}(q|y)=p^C_Q(q)$. Define $p_{Q'|XY}$ over the set
of connected components in this graph
such that,
with probability 1, $Q'$ is the connected
component $C(X,Y)$ in this graph to which the vertices $X$ and $Y$ belong
(and hence $H(Q'|X)=H(Q'|Y)=0$), and 
$p_{Q|Q'}(q|C)=p^C_Q(q)$. Then
$p_{Q|XY}(q|x,y)=p_{Q|X}(q|x)=p^{C(x,y)}_Q(q)=p_{Q|Q'}(q|C(x,y))$, so that
$Q-Q'-XY$.
\end{IEEEproof}

\medskip

The following calculation is useful in applying the above lemma in a couple
of our proofs.
\begin{lem}
\label{lem:connectedcomp-addon}
For correlated random variables $(X,Y,Q,Q')$
if $H(Q'|X)=0$ (or $H(Q'|Y)=0$) and $Q-Q'-XY$,
then $I(X;Y|Q) \ge I(X;Y|Q')$.
\end{lem}
\begin{IEEEproof}
\begin{align*}
I(X;Y|Q) &\stackrel{\text{(a)}}{=} I(X,Q';Y|Q)\\
	&\geq I(X;Y|Q',Q)\\
	&\stackrel{\text{(b)}}{=} I(X;Y|Q'),
\end{align*}
where the (a) follows from $H(Q'|X)=0$ and (b) from the Markov chain $Q - Q' - XY$. 
\end{IEEEproof}

\medskip
\begin{IEEEproof}[Proof of \theoremref{intercepts}]
To prove \eqref{eq:Tintx}, 
firstly note that $ \Tintx XY = $
\[\inf_{\substack{\pQXY:\\I(X;Q|Y)=0\\I(X;Y|Q)=0}} I(Y;Q|X) \le
\inf_{\substack{\pQXY:\\H(Q|Y)=0\\I(X;Y|Q)=0}} H(Q|X),\]
because if $H(Q|Y)=0$ then $I(X;Q|Y)=0$ and $I(Y;Q|X)=H(Q|X)$. For the other direction, we invoke
\lemmaref{connectedcomp} (with $X$ and $Q$ interchanged), so that given $Q$
such that $I(X;Q|Y)=I(X;Y|Q)=0$, $\exists Q'$ such that $H(Q'|Y)=H(Q'|Q)=0$
and $X-Q'-QY$; then $H(Q'|X) = I(Y;Q'|X)\le I(Y;Q|X)$, and $X-Q'-Y$. So $Q'$
is considered in the $\inf$ expression of the RHS, and we have LHS $\ge$ RHS.
This proves \eqref{eq:Tintx}.
Similarly, \eqref{eq:Tinty} holds.  

To prove \eqref{eq:Tintz}, firstly we note that $\Tintz XY =$
\[\inf_{\substack{\pQXY:\\I(Y;Q|X)=0\\I(X;Q|Y)=0}} I(X;Y|Q) \le \inf_{\substack{\pQXY:\\H(Q|X)=0\\H(Q|Y)=0}} I(X;Y|Q),\]
since $H(Q|X) = H(Q|Y) = 0$ implies that $I(Y;Q|X) = I(X;Q|Y) = 0$. 
For the inequality in the other direction, 
by \lemmaref{connectedcomp}, given 
$Q$ such that $I(Y;Q|X)=I(X;Q|Y)=0$, we get $Q'$ such that 
$H(Q'|X)=H(Q'|Y)=0$ and $Q-Q'-XY$; then, by \lemmaref{connectedcomp-addon}
it follows that $I(X;Y|Q) \ge I(X;Y|Q')$.
Hence,
$\inf_{\pQXY:I(Y;Q|X)=I(X;Q|Y)=0} I(X;Y|Q) \ge \inf_{\pQXY:H(Q|X)=H(Q|Y)=0} I(X;Y|Q)$.
Thus, \eqref{eq:Tintz} holds.  
\end{IEEEproof}

\medskip
\begin{IEEEproof}[Proof of \theoremref{Rtens-convex}]
Consider any two points $s_1,s_2\in\Rtens XY$. 
Consider any point $s=\alpha s_1 +
(1-\alpha) s_2$ for $0\le\alpha\le1$. We need to show that $s\in\Rtens XY$
as well.

Since $s_1,s_2\in\Rtens XY$, there are
random variables $p_{Q_1|XY}$ and $p_{Q_2|XY}$ such that
$s_1' := \tens XY{Q_1} \le s_1$ and 
$s_2' := \tens XY{Q_2} \le s_2$. 
Let $J$ be a binary random variable independent of $(X,Y,Q_1,Q_2)$ taking
on value 1 with probability $\alpha$ and 2 with probability $1-\alpha$. Let
$Q=(J,Q_J)$.
Then $\tens XYQ=\alpha \tens XY{Q_1} + (1-\alpha) \tens
XY{Q_2}$. That is, $s'=\alpha s'_1 + (1-\alpha)s'_2$
is in \Rtens XY. Hence $s\in\Rtens XY$, since $s\ge s'$.
\end{IEEEproof}

\begin{lem}\label{lem:closedness-of-ihull}
If $A\subseteq \Real^m$ is compact, then its increasing hull,
\[ \ihull{A}=\{x\in\Real^m: x\geq a \text{ for some } a\in A\},\]
is closed.
\end{lem}
\begin{IEEEproof}
Let $\{x_n\}$ be a sequence in $\ihull{A}$ converging to $x$. Then, there is a sequence $\{a_n\}$ in $A$ such that $x_n \geq a_n$, for all $n$. Since $A$ is compact, there is a convergent subsequence $\{a_{n_k}\}$ of $\{a_n\}$ that converges to $a\in A$. Also, the subsequence $\{x_{n_k}\}$ converges to $x$, and satisfies $x_{n_k} \geq a_{n_k}$, for all $k$. Thus, $x\geq \lim_{k\rightarrow\infty} a_{n_k} =a$, and so, $x\in \ihull{A}$.
\end{IEEEproof}

The following simple (and standard) observation is used
in proving \lemmaref{statdiff-mutualinfo}.
\begin{lem}
\label{lem:statdiff-coupling}
If $p_Z$ and $p_{Z'}$ are such that $\Delta(Z,Z')=\epsilon$,
then there is a joint distribution
$p_{JWW'}$ such that $p_W=p_Z$, $p_{W'}=p_{Z'}$, $p_J(0)=\epsilon$ and
$p_J(1)=1-\epsilon$ and $J=1\implies W=W'$.
\end{lem}
\begin{IEEEproof}
First we define independent random variables $J$, $W_0$, $W_1$ and $W_2$
(the first one over $\{0,1\}$ and the others over the common alphabet of 
$Z$ and $Z'$ as follows.
\begin{align*}
p_J(0)&=\epsilon, \text{ and } p_J(1)=1-\epsilon,\\
p_{W_0}(z)&=\frac{\min\{p_{Z}(z),p_{Z'}(z)\}}{1-\epsilon},\\
p_{W_1}(z)&= \frac{p_{Z}(z)- (1-\epsilon) \cdot p_{W_0}(z)}{\epsilon},\\
p_{W_2}(z)&= \frac{p_{Z'}(z)- (1-\epsilon) \cdot p_{W_0}(z)}{\epsilon}.
\end{align*}
We define $W$ and $W'$ in terms of these random variables:
when $J=1$, $W=W'=W_0$, and when $J=0$ we set $W=W_1$ and $W'=W_2$.
It is easy to verify that the resulting random variables have the correct
marginals.
\end{IEEEproof}

\paragraph*{\lemmaref{statdiff-mutualinfo}}
Suppose random variables $(A,B,C)$ and $(A',B',C')$ over the same alphabet
$\sA\times\sB\times\sC$
are such that $\Delta(ABC,A'B'C')=\epsilon$.  Then $I(A';B'|C') \le
I(A;B|C) + 2H_2(\epsilon) + \epsilon\log\min\{|\sA|,|\sB|\}$.

\begin{IEEEproof}
We apply \lemmaref{statdiff-coupling} with $Z=(A,B,C)$ and
$Z'=(A',B',C')$ to obtain a joint distribution 
$p_{J,A,B,C,A',B',C'}$ so that $J=1\implies (A,B,C)=(A',B',C')$
and this event occurs with probability $1-\epsilon$.

Now, note that 
\begin{align*}
I(A;B|C) &= I(A;BJ|C) - I(A;J|BC) \\ &= I(A;B|CJ) + I(A;J|C)-I(A;J|BC).
\end{align*}
Since $0 \le I(A;J|C)\le H(J)$ and $0\le I(A;J|BC) \le H(J)$,
we have
\begin{align}
\label{eq:statdiff-mutualinfo-bound}
|I(A;B|C) - I(A;B|CJ)| \le H(J) =  H_2(\epsilon)
\end{align}

The same condition holds for $A',B',C'$ instead of $A,B,C$. Hence
\begin{align*}
& I(A';B'|C') \le I(A';B'|C'J) + H_2(\epsilon) \\
&\qquad = (1-\epsilon)I(A';B'|C',J=1)\\*
&\qquad\qquad\qquad\qquad + \epsilon I(A';B'|C',J=0) + H_2(\epsilon)\\
&\qquad = (1-\epsilon)I(A;B|C,J=1)\\*
&\qquad\qquad\qquad\qquad + \epsilon I(A';B'|C',J=0) + H_2(\epsilon)\\
&\qquad = I(A;B|CJ) - \epsilon I(A;B|C,J=0) \\*
&\qquad\qquad\qquad\qquad + \epsilon I(A';B'|C',J=0) + H_2(\epsilon)\\
&\qquad\namedleq{a} I(A;B|C) + \epsilon I(A';B'|C',J=0) + 2H_2(\epsilon)\\
&\qquad\leq I(A;B|C) + \epsilon\min\{\log |\sA|,\log |\sB|\} + 2H_2(\epsilon),
\end{align*}
where (a) follows from \eqref{eq:statdiff-mutualinfo-bound}.
\end{IEEEproof}

\section{Details Omitted from \sectionref{ACI}}
\label{app:ACI}

\medskip
\begin{IEEEproof}[Proof of \corollaryref{GKW-tension}]
The first equation \eqref{eq:Rtens-ARI} follows immediately from
\theoremref{GKW-char}. We need to show \eqref{eq:Rtens-ACI} which is
repeated below for convenience.
\begin{align*}
\Rtens XY &= \ihull{f_{X,Y}(\sRsACI(X;Y))}  &&&\qquad\eqref{eq:Rtens-ACI}
\end{align*}
where $f_{X,Y}$ is an affine map defined as
\begin{align*}
f_{X,Y}\left(\left[\begin{array}{c}R_1\\R_2\\R_\CI\end{array}\right]\right)
\defineqq
 \left[\begin{array}{c}R_1\\ R_2\\I(X;Y) + R_1+R_2 -R_\CI\end{array} \right].
\end{align*}
Given a $p_{Q|XY}$ and $(r_1,r_2,r_\CI)$ such that $r_1\geq I(Y;Q|X)$,
$r_2\geq I(X;Q|Y)$ and $r_\CI\leq I(XY;Q)$, we have
\begin{align*}
r_1&+r_2-r_\CI+I(X;Y)\\ &\geq I(Y;Q|X)+I(X;Q|Y) - I(XY;Q) + I(X;Y)\\
                     &= I(X;Y|Q),
\end{align*}
where the last equality is~\eqref{eq:MIeqs4}.
Thus, $\mathrm{L.H.S.}\supseteq\mathrm{R.H.S.}$

If $(r'_1,r'_2,r'_3)\in\Rtens XY$, then there is a $p_{Q|XY}$ such that
$r'_1\geq I(Y;Q|X)$, $r'_2\geq I(X;Q|Y)$ and $r'_3\geq I(X;Y|Q)$. But,
since \eqref{eq:MIeqs4}
implies that $(I(Y;Q|X),I(X;Q|Y),I(X;Y|Q))\in f_{X,Y}(\sRsACI(X;Y))$, 
we have $(r'_1,r'_2,r'_3)\in \ihull{f_{X,Y}(\sRsACI(X;Y))}$. Thus,
$\mathrm{L.H.S.}\subseteq\mathrm{R.H.S.}$

\end{IEEEproof}

\medskip
\begin{IEEEproof}[Proof of \corollaryref{GacsKo}]
From the definitions it is clear that, $\Com \GKW(X;Y) \leq \RACIintz XY$.
But as we will show, this is in fact an equality. \theoremref{GKW-char}
implies that
\begin{align}
\RACIintz XY&= \max_{\substack{p_{Q|XY}:\\I(X;Q|Y)=I(Y;Q|X)=0}} I(XY;Q).
\label{eq:RACIintz-char}
\end{align}
By \lemmaref{connectedcomp}, given \pQXY such that $I(X;Q|Y)=I(Y;Q|X)=0$,
we can find a random variable $Q'$ with $H(Q'|X)=H(Q'|Y)=0$ and
$Q-Q'-(X,Y)$ is a Markov chain. Then, clearly, $I(X;Q'|Y) = I(Y;Q'|X) = 0$
and furthermore
\begin{align*} 
I(XY;Q)\leq I(XY;QQ') = I(XY;Q') = H(Q').
\end{align*}
Hence,
\begin{align*}
\RACIintz XY&= \max_{p_{Q'|XY}:H(Q'|X)=H(Q'|Y)=0} H(Q').
\end{align*}
Since $H(Q'|X)=H(Q'|Y)=0$, $Q'=f_1(X)$ and $Q'=f_2(Y)$
for some functions $f_1$ and $f_2$, and hence
$\Com \GKW(X;Y) \geq H(Q')$. So, $\Com\GKW(X;Y) \geq \RACIintz XY$. Hence,
we can conclude \eqref{eq:CGKWequivalence}-\eqref{eq:CGKW}. 

It only remains to show
\begin{align}
 \Tintz XY = I(X;Y) - \RACIintz XY.\label{eq:CGKWprimefromRIO}
\end{align}
This easily follows from \eqref{eq:Tint} and \eqref{eq:RACIintz-char}
using \eqref{eq:MIeqs1}-\eqref{eq:MIeqs3}.
\end{IEEEproof}

\medskip
\begin{IEEEproof}[Proof of \lemmaref{achievabilitylemma}]

We are given $p^\phast_{X,Y}$, $p^\ast_{Q|XY}$, $d$. Also, we have \[
D^\ast = \Exp[p^\phast_{X,Y}p^\ast_{Q|XY}]{d(X,Y,Q)}.\] This proof uses the
notion of typicality. We will use notation, definitions, and results
from~\cite{ElGamalKim2012}. All typical sequences are defined with respect
to the joint distribution $p^\phast_{X,Y}p^\ast_{Q|XY}$. For a positive
integer $k$, we will denote $\{1,\ldots,k\}$ by $[k]$.

\noindent{\em Random codebook construction:} Let $\epsilon'>0$
and $p_Q(q)=\sum_{x,y} p^\phast_{X,Y}(x,y)p^\ast_{Q|XY}(q|x,y)$ be the
marginal distribution of $Q$ induced by the given joint distribution. Let
$r,r_1,r_2$ be such that $r\geq r_1,r_2$. We generate $2^{nr}$ codewords
$Q^n(l), l\in [2^{nr}]$ randomly and independently each according to
$\prod_{i=1}^n p_Q(q_i)$. The set of indices $l\in [2^{nr}]$ is then
partitioned in two different ways into equal size subsets: {\em 1-bins}
$\sB_1(m_1) = \{ (m_1-1)2^{n(r-r_1)}+1, \ldots, m_1 2^{n(r-r_1)} \}, m_1\in
[2^{nr_1}]$, and {\em 2-bins} $\sB_2(m_2) = \{ (m_2-1)2^{n(r-r_2)}+1,
\ldots, m_2 2^{n(r-r_2)} \}, m_2 \in [2^{nr_2}]$.

\noindent{\em Encoding:} If the input to the encoder is $(x^n,y^n)$, it
finds an index $l$ such that $(x^n,y^n,q^n(l))\in
\typical[X,Y,Q]{\epsilon'}$. If none is available, $l$ is chosen uniformly
at random from $[2^{nr}]$. The encoder sends to the $k$-th receiver,
$k=1,2$, the bin index $m_k$ such that $l\in\sB_k(m_k)$, i.e.,
$f\upn_k(x^n,y^n)=m_k$, $k=1,2$.

\noindent{\em Decoding:} The first decoder, on receiving $m_1$, tries to
find a unique $\hl_1\in \sB_1(m_1)$ such that $(x^n,q^n(\hl_1))\in
\typical[X,Q]{\epsilon'}$. If it cannot find such an $\hl_1$, it sets
$\hl_1=1$. Decoder~1 outputs $\hl_1$, i.e., $g\upn_1(x^n,m_1)=\hl_1$.
Similarly, decoder~2 outputs a $\hl_2$ it finds using $y^n,m_2$, and
$\sB_2$.  

\noindent{\em Reconstruction:} The reconstruction function $h\upn$ is
defined as $h\upn(l)=q^n(l)$. Thus the output sequence is
\[ q^n = h\upn(\hl_1)=q^n(\hl_1).\]

\noindent{\em Analysis of the probability of error and expected
distortion:} Let $L,M_1,M_2,\hL_1,\hL_2$ be the indices chosen by the
encoder and the decoder. We define the {\em error event} as
\begin{align*}
 &\sE =\\ &\;\left\{ \hL_1\neq\hL_2 \right\}
     \cup \left\{ (X^n,Y^n,Q^n(\hL_1)) \notin \typical[X,Y,Q]{\epsilon'}
\right\}.
\end{align*}
Let
\begin{align*}
\sE_0 &= \left\{ (X^n,Y^n,Q^n(l))\notin \typical{\epsilon'}
\text{ for all } l\in [2^{nr}] \right\},\\
\sE_1 &= \{ (X^n,Q^n(\tl_1)) \in \typical{\epsilon'}\text{ for some } \\
&\qquad\qquad\qquad\qquad\qquad\qquad \tl_1\in \sB_1(M_1), \tl_1\neq L\},\\
\sE_2 &= \{ (Y^n,Q^n(\tl_2)) \in \typical{\epsilon'}\text{ for some } \\
&\qquad\qquad\qquad\qquad\qquad\qquad \tl_2\in \sB_2(M_2), \tl_2\neq L\}.
\end{align*}
Since the error event occurs only when $(X^n,Y^n,Q^n(L))\notin
\typical{\epsilon'}$ or at least one of $L_1$ and $L_2$ is different from
$L$, we have
\[ \sE \subseteq \sE_0\cup\sE_1\cup\sE_2. \]
By union bound,
\[ \Pr(\sE) \leq \Pr(\sE_0) + \Pr(\sE_1) + \Pr(\sE_2).\]
By covering lemma~\cite[Lemma 3.3]{ElGamalKim2012}, $\Pr(\sE_0) \rightarrow
0$ as $n\rightarrow \infty$ provided $r > I(X,Y;Q) + \delta(\epsilon')$,
where $\delta(\epsilon')\downarrow 0$ as $\epsilon'\downarrow 0$. To
upperbound $\Pr(\sE_1)$, we claim that
\begin{align*}
 &\Pr(\sE_1) \leq \\&\quad\Pr\left(\left\{ (X^n,Q^n(\tl_1))\in \typical{\epsilon}
\text{ for some } \tl_1\in \sB_1(1) \right\} \right). 
\end{align*}
For a proof see~\cite[Lemma 11.1, pg. 284]{ElGamalKim2012}. For each
$\tl_1\in\sB_1(1)$, the codeword $Q^n(\tl_1)$ is generated independent of
$X^n$ and according to $\prod_{i=1}^n p_Q(q_i)$. Note that there are
$2^{n(r-r_1)}$ codewords in $\sB_1(1)$. By packing lemma~\cite[Lemma
3.1]{ElGamalKim2012}, the probability term on the R.H.S. above tends to
zero as $n \rightarrow \infty$ provided $r - r_1 \leq I(X;Q) -
\delta(\epsilon')$. Similarly, $\Pr(\sE_2)\rightarrow 0$ as $n\rightarrow
\infty$ if $r-r_2 \leq I(Y;Q) - \delta(\epsilon')$. Combining the
conditions for all three events, we have $\Pr(\sE)\rightarrow 0$ as $n
\rightarrow \infty$ provided
\begin{align}
\begin{split}
r_1 &\geq I(Y;Q|X) + 2\delta(\epsilon'),\\
r_2 &\geq I(X;Q|Y) + 2\delta(\epsilon').
\end{split} \label{eq:rateconds}
\end{align}
We have shown that, when \eqref{eq:rateconds} hold, the ensemble average of
$\Pr(\sE)$ over $(2^{nr},2^{nr_1},2^{nr_2},n)$ codes converges to zero as
$n\rightarrow 0$. Hence, we can assert that there must exist a sequence of
(deterministic) $(2^{nr},2^{nr_1},2^{nr_2},n)$ codes such that $\Pr(\sE)
\rightarrow 0$ as $n \rightarrow \infty$ if \eqref{eq:rateconds} is
satisfied. Clearly, with an appropriately small choice of $\epsilon'$, this
sequence of codes satisfies the rate conditions \eqref{eq:ratescondition}
with $R_1=I(Y;U|X)$ and $R_2=I(X;U|Y)$, and also the probability of error
condition \eqref{eq:Pecondition}. It only remains to verify
\eqref{eq:dist} which we do below:
\begin{align*}
 \frac{1}{n}&\sum_{i=1}^n\Exp{d(X_i,Y_i,Q_i)}\\
   &\leq d_\mathrm{max} \Pr(\sE) + \Exp{\frac{1}{n} \sum_{i=1}^n d(X_i,Y_i,Q_i)
\middle|\sE^c}\\
  &\leq d_\mathrm{max} \Pr(\sE) + (1+\epsilon')\Exp{d(X,Y,Q)},
\end{align*}
where the last inequality follows from the typical average
lemma~\cite[pg. 26]{ElGamalKim2012}. Thus, for a small enough choice of
$\epsilon'$, we can satisfy \eqref{eq:dist} as well with $D=D^\ast$.
\end{IEEEproof}

\section{Details Omitted from \sectionref{GrayWyner}}
\label{app:GrayWyner}

\medskip
\begin{IEEEproof}[Proof of \theoremref{affine}]

It is easy to prove this theorem from the definition of $\Rtens{X}{Y}$ (\definitionref{tension})
and~\theoremref{GW} by making use of the mutual information equalities
\eqref{eq:MIeqs1}-\eqref{eq:MIeqs3} at the top of
page~\pageref{float:affine}.

\end{IEEEproof}

\medskip
\begin{IEEEproof}[Proof of \corollaryref{CGKWfromThm}]
\begin{align*}
&\sup\{R_\C:R_\A+R_\C=H(X),\\
 &\qquad\qquad 
  R_\B+R_\C=H(Y), (R_\A,R_\B,R_\C)\in\sRsGW\}\\
&\namedeqq{a}\sup\{R:(0,0,I(X;Y)-R)\in\sRsGW'\}\\
&\namedeqq{b}\sup\{R:(0,0,I(X;Y)-R)\in\Rtens XY\}\\
&\namedeqq{c}\Com \GKW(X;Y),
\end{align*}
where (a) follows from the definition $\sRsGW'=f(\sRsGW)$. The $\leq$
direction of (b) follows directly from \theoremref{affine}. But $<$ cannot
hold since if $(0,0,I(X;Y)-R)\in\Rtens XY$, then there is a $R'\geq R$ such
that $(0,0,I(X;Y)-R')\in\sRsGW'$. Finally, (c) follows from
\corollaryref{GacsKo}.

To arrive at the alternative form, we verify the equivalence of the two
forms.
\begin{align*}
&\{R: R \leq I(X;Y),%\notag\\ &\qquad\qquad\;\;
   \{R_\C=R\} \cap \GWlower \subseteq \sRsGW\}\\
&\,= \{R_\C: R_\A+R_\C=H(X),\\
   &\qquad\qquad\,
  R_\B+R_\C=H(Y), (R_\A,R_\B,R_\C)\in \sRsGW\}.
\end{align*}
$\subseteq$: if $R \leq I(X;Y)$, then $(H(X)-R,H(Y)-R,R)\in
\{R_\C=R\}\cap\GWlower$.\\
$\supseteq$: Let $s=(H(X)-R_\C,H(Y)-R_\C,R_\C)\in\sRsGW$. Then (a) $R_C\leq
I(X;Y)$ since $s\in\GWlower$, and (b) if $s'=(r_\A,r_\B,R_\C)\in\GWlower$,
then since $r_\A \geq H(X)-R_\C$ and $r_\B\geq H(Y)-R_\C$, we have $s'\geq
s$ (component-wise) which implies that $s'\in\sRsGW$ from the
definition of the \gw system.
\end{IEEEproof}

\medskip
\begin{IEEEproof}[Proof of \corollaryref{CGWfromThm}]
\begin{align*}
\Com\Wyner &= \inf \{R_\C: (R_\A,R_\B,R_\C)\in \sRsGW,\\
       &\qquad\qquad\qquad R_\A+R_\B+R_\C=H(X,Y)\}\\
  &\namedeqq{a} \inf \{R_1+R_2+I(X;Y): (R_1,R_2,0)\in\sRsGW'\}\\
  &\namedeqq{b} \inf \{R_1+R_2+I(X;Y): (R_1,R_2,0)\in\Rtens XY\},
\end{align*}
where (a) follows from the definition $\sRsGW'=f(\sRsGW)$; (b) follows from
\theoremref{affine}: $\geq$ direction follows directly from the theorem.
But $>$ cannot hold, since by the theorem, if $(R_1,R_2,0)\in\Rtens XY$ then
there exists $(R_1',R_2',0)\in\sRsGW'$ such that $R_1'\leq R_1$ and
$R_2'\leq R_2$.

\end{IEEEproof}

\medskip
\begin{IEEEproof}[Proof of \corollaryref{cornerconnection}]
\begin{align*}
&G(Y\rightarrow X)\\
&\quad=\inf\{R_\C:(H(X|Y),H(Y)-R_\C,R_\C) \in\sRsGW\},\\
&\quad\namedeqq{a}\inf\{R:(R-I(X;Y),0,0)\in\sRsGW'\}\\
&\quad\namedeqq{b}\inf\{R:(R-I(X;Y),0,0)\in\Rtens XY\}\\
&\quad\namedeqq{c}I(X;Y)+\Tintx XY,
\end{align*}
where (a) follows from $\sRsGW'=f(\sRsGW)$. (b) is a consequence of
\theoremref{affine}: And (c) follows from the definition of $\Tintx XY$.

Similarly we get \eqref{eq:cornerresult2}. The equality
\eqref{eq:kamathdual} is proved in~\cite{KamathAn10} which along
with \eqref{eq:cornerresult1}-\eqref{eq:cornerresult2} implies
\eqref{eq:kamathdualresult}.

\end{IEEEproof}

\section{Details Omitted from \sectionref{crypto}}
\label{app:crypto}

Here we prove \theoremref{robust-monotone-stat-rate-bound}. 
The following lemma will be useful in this.
\begin{lem}
\label{lem:simulation-to-smalldependency}
Suppose $\statsamples\epsilon\Pi{(X,Y)}{(U,V)}$. Then
\begin{align*}
I(\Pialice;\Pibobout|\Pialiceout) &\le 2\delta(\epsilon) \\
I(\Pibob;\Pialiceout|\Pibobout)   &\le 2\delta(\epsilon)
\end{align*}
where $\delta(\epsilon) = 2H_2(\epsilon) + \epsilon\log\max\{|\sU|, |\sV|\}$.
\end{lem}
\begin{IEEEproof}
We show $I(\Pialice;\Pibobout|\Pialiceout)\le2\delta(\epsilon)$ 
(the other relation following similarly).
Let \simPialice be as in \definitionref{stat-secure-sampling}.
Then $I(\simPialice;V|U)=0$ and $\Delta(\simPialice V, \Pialice \Pibobout)
\le \epsilon$. Also, we have $\Delta(UV,\Pialiceout\Pibobout)\le\epsilon$.  Then
\begin{align*}
&I(\Pialice;\Pibobout|\Pialiceout) \\
&\qquad = I(\Pialice;\Pibobout|\Pialiceout)-I(\simPialice;V|U) \\
&\qquad \namedeqq{a}
   \Big[ H(\Pibobout|\Pialiceout) - H(V|U) \Big] \\*
&\qquad\qquad\qquad - H(\Pibobout|\Pialice) + H(V|U\simPialice) \\
&\qquad = \Big[ H(\Pibobout|\Pialiceout) - H(V|U) \Big] \\*
&\qquad\qquad\qquad + \Big[ H(V|\simPialice) - H(\Pibobout|\Pialice) \Big] \\*
&\qquad\qquad\qquad - I(V;U|\simPialice) \\
&\qquad \namedleq{b}
   2\delta(\epsilon)
\end{align*}
where in (a) we used
$H(\Pibobout|\Pialice\Pialiceout)=H(\Pibobout|\Pialice)$ (because
\Pialiceout is a function of \Pialice) and in (b) we 
bounded the two terms in the square brackets by
invoking \lemmaref{statdiff-mutualinfo} twice,
with $((ABC),(A'B'C'))$ being $((VVU),(\Pibobout\Pibobout\Pialiceout))$
and $((VV\simPialice),(\Pibobout\Pibobout\Pialice))$ respectively.
\end{IEEEproof}

\medskip
\begin{IEEEproof}[Proof of \theoremref{robust-monotone-stat-rate-bound}]
Suppose there is a protocol $\Pi$ such that
$\statsamples\epsilon\Pi{(X^{n_2},Y^{n_2})}{(U^{n_1},V^{n_1})}$,
for $\frac{n_1}{n_2}\ge r-\epsilon'$.
We will denote the final
views of the two parties in this protocol by $(\Pialice,\Pibob)$. Also, we shall
denote the outputs by $(\Pialiceout,\Pibobout)$. Then, firstly,
by conditions (1) and (2) of \definitionref{monotone},
\[ \M {\Pialice} {\Pibob} \supseteq \M {X^{n_2}}{Y^{n_2}}.\]

Secondly, by \lemmaref{simulation-to-smalldependency},
for random variables $(\Pialice,\Pialiceout,\Pibobout,\Pibob)$,
the hypothesis in condition~(3$'$)  
of \definitionref{robust-monotone} holds,
with $\phi=\Del(\epsilon) \cdot n_1 \cdot \log|\sU||\sV|$
where we set  
$\Del(\epsilon)=2(2H_2(\epsilon)+\epsilon)$.
Hence 
\begin{align*}
\M {\Pialiceout} {\Pibobout} &\supseteq \M {\Pialice} {\Pibob} \\*
&\qquad\qquad + c\Del(\epsilon) \cdot n_1 \log|\sU||\sV|,
\end{align*}
where $c$ is as in \definitionref{robust-monotone}.
Finally, since $\Delta(U^{n_1}V^{n_1},\Pialiceout\Pibobout)\le\epsilon$, by
the continuity of \Mfunc
(condition~(3$''$) of \definitionref{robust-monotone}), 
we have
\begin{align*}
\M {U^{n_1}} {V^{n_1}} &\supseteq \M {\Pialiceout} {\Pibobout} \\*
&\qquad\qquad + \del(\epsilon)\cdot n_1 \log|\sU||\sV|,
\end{align*}
where $\del(\epsilon)$ is as in 
condition~(3$''$) of \definitionref{robust-monotone}. 
Putting these together, after dividing throughout by $n_1$ (using condition (4) in
\definitionref{monotone} and convexity from condition~(3$''$)), and using $\frac{n_2}{n_1} \le \frac1{r-\epsilon'}$, we get
\[  \M UV \supseteq \frac{1}{r-\epsilon'} \M XY + \del'(\epsilon) \cdot \log|\sU||\sV|,\]
where $\del'(\epsilon)=c\Del(\epsilon)+\del(\epsilon)$.

If the rate of statistically securely sampling $(U,V)$ from $(X,Y)$ is $r$,
then for all $\epsilon,\epsilon'>0$, the above relation should hold. Since
$\del'(\epsilon)\downarrow 0$ as $\epsilon \downarrow 0$ and the regions 
\M UV and \M XY are closed (condition (3$''$)), we get
\[  \M UV \supseteq \frac1r \M XY \]
as required.
\end{IEEEproof}


\begin{thebibliography}{1}

\bibitem{AhlswedeKo74}
R. Ahlswede and  J. K\"{o}rner,
\newblock ``On common information and related characteristics of correlated
information sources,''
\newblock in {\em Proc. of the 7th Prague Conference on Information
Theory}, 1974.

\bibitem{Beaver96}
D. Beaver,
\newblock ``Correlated pseudorandomness and the complexity of private
  computations,''
\newblock in {\em Proc.\ $28$th STOC}, pp. 479--488, ACM, 1996.

\bibitem{Beaver95}
D. Beaver,
\newblock ``Precomputing oblivious transfer,''
\newblock in Don Coppersmith, editor, {\em CRYPTO}, vol. 963 of {\em Lecture
  Notes in Computer Science}, pp. 97--109, Springer, 1995.

\bibitem{CoverT06}
T.~M. Cover and J.~A. Thomas, {\em Elements of Information Theory}, 2ed,
Wiley, 2006.

\bibitem{CsiszarAh07}
I. Csisz\'{a}r and R. Ahlswede,
\newblock ``On oblivious transfer capacity,''
\newblock in {\em Proc. International Symposium on Information Theory
(ISIT)}, pp. 2061--2064, 2007.

\bibitem{CsiszarKo81}
I. Csisz\'ar and J. K\"orner,
\newblock {\em Information Theory: Coding Theorems for Discrete Memorless Systems}, 1ed, Akad\'emiai Kiad\'o, Budapest, 1981.

\bibitem{DodisMi99}
Y. Dodis and S. Micali,
\newblock ``Lower bounds for oblivious transfer reductions,''
\newblock in Jacques Stern, editor, {\em EUROCRYPT}, vol. 1592 of {\em
  Lecture Notes in Computer Science}, pp. 42--55, Springer, 1999.

\bibitem{ElGamalKim2012}
A. El Gamal and Y.-H. Kim,
\newblock {\em Network Information Theory}, Cambridge, 2012.

\bibitem{GacsKo73}
P. G\'{a}cs and J. K\"{o}rner,
\newblock ``Common information is far less than mutual information,''
\newblock {\em Problems of Control and Information Theory}, vol. 2, no. 2, pp. 119--162,
 1973.

\bibitem{GastparRV03}
M. Gastpar, B. Rimoldi, and M. Vetterli, ``To code or not to code: Lossy
source-channel communication revisited,''
{\em IEEE Transactions on Information Theory}, vol. 49, no. 5, pp.
1147--1158, 2003.

\bibitem{GrayWy74}
R. M. Gray and A. D. Wyner, ``Source coding for a simple network,'' 
{\em Bell System Technical Journal}, vol. 53, no. 9, pp. 1681–1721, 1974.

\bibitem{ImaiMoNa06}
H. Imai, K. Morozov, and A. C.~A. Nascimento,
\newblock ``On the oblivious transfer capacity of the erasure channel,''
\newblock in {\em Proc. International Symposium on Information Theory
(ISIT)}, pp. 1428--1431, 2006.

\bibitem{ImaiMoNa07}
H. Imai, K. Morozov, and A.~C.~A. Nascimento,
\newblock ``Efficient oblivious transfer protocols achieving a non-zero
rate from any non-trivial noisy correlation,''
\newblock in {\em Proc. International Conference on Information Theoretic Security
(ICITS)}, 2007.

\bibitem{ImaiMoNaWi06}
H. Imai, K. Morozov, A.~C.~A. Nascimento, and A. Winter,
\newblock ``Efficient protocols achieving the commitment capacity of noisy
  correlations,''
\newblock in {\em Proc. International Symposium on Information Theory (ISIT)}, pp.
 1432--1436, 2006.

\bibitem{ImaiMuNaWi04}
H. Imai, J. M\"uller-Quade, A.~C.~A. Nascimento, and A. Winter,
\newblock ``Rates for bit commitment and coin tossing from noisy
correlation,''
\newblock in {\em Proc. International Symposium on Information Theory (ISIT)},
pp. 45, 2004.

\bibitem{IshaiKuOsSa09}
Y. Ishai, E. Kushilevitz, R. Ostrovsky, and A. Sahai,
\newblock ``Extracting Correlations.''
\newblock  in {\em Proc.\ $50$th FOCS}, pp. 261--270, IEEE, 2009.

\bibitem{KamathAn10}
S. Kamath and V. Anantharam, ``A new dual to the G\'acs-K\"orner common
information defined via the Gray-Wyner system,'' in {\em Proc. 48th Allerton
Conf. on Communication, Control, and Computing}, pp. 1340--1346, 2010.

\bibitem{Kilian88}
J. Kilian,
\newblock ``Founding cryptography on oblivious transfer,''
\newblock in {\em Proc.\ $20$th STOC}, pp. 20--31, ACM, 1988.

\bibitem{Kilian00}
J. Kilian,
\newblock ``More general completeness theorems for secure two-party
computation,''
\newblock in {\em Proc.\ $32$nd STOC}, pp. 316--324, ACM, 2000.

\bibitem{MarcoEf09}
D. Marco and M. Effros, ``On lossless coding with coded side information,''
{\em IEEE Transactions on Information Theory}, vol. 55, no. 7, pp. 3284--3296, 2009.

\bibitem{PrabhakaranPr10}
V. M. Prabhakaran and M. M. Prabhakaran,
\newblock ``Assisted common information,''
\newblock in {\em Proc. International Symposium on Information Theory (ISIT)},
pp. 2602-2606, 2010. 

\bibitem{PrabhakaranPr11}
V. M. Prabhakaran and M. M. Prabhakaran,
\newblock ``Assisted common information: Further results,''
\newblock in {\em Proc. International Symposium on Information Theory (ISIT)},
pp. 2861 - 2865, 2011. 

\bibitem{PrabhakaranPr12}
M. M. Prabhakaran and V. M. Prabhakaran,
\newblock ``On secure multiparty sampling for more than two parties,''
\newblock in {Proc. IEEE Information Theory Workshop (ITW)}, pp. 99 - 103, 2012.

\bibitem{Rabin81}
M.~Rabin.
\newblock ``How to exchange secrets by oblivious transfer,''
\newblock Technical Report TR-81, Harvard Aiken Computation Laboratory, 1981.

\bibitem{Wiesner83}
Stephen Wiesner.
\newblock ``Conjugate coding,''
\newblock Sigact News, vol. 15, pp.~78--88, 1983.

\bibitem{WinklerWu10}
S. Winkler and J. Wullschleger.
\newblock ``On the Efficiency of Classical and Quantum Oblivious Transfer Reductions,''
\newblock in Tal Rabin, editor, {\em CRYPTO}, vol. 6223 of {\em Lecture
  Notes in Computer Science}, pp. 707--723, Springer, 2010.

\bibitem{WinterNaIm03}
A. Winter, A. C.~A. Nascimento, and H. Imai.
\newblock ``Commitment capacity of discrete memoryless channels,''
\newblock in Kenneth~G. Paterson, editor, {\em IMA Int. Conf.}, vol. 2898 of
  {\em Lecture Notes in Computer Science}, pp. 35--51, Springer, 2003.

\bibitem{Witsenhausen75}
H. S. Witsenhausen, 
\newblock ``On sequences of pairs of dependent random variables,''
\newblock {\em SIAM Journal of Applied Mathematics},  28:100--113, 1975.

\bibitem{WolfWu06}
S. Wolf and J. Wullschleger.
\newblock ``Oblivious Transfer Is Symmetric,''
\newblock {in Serge Vaudenay, editor,} {\em EUROCRYPT}, vol. 4004 of {\em Lecture
  Notes in Computer Science}, pp. 222--232, Springer, 2006.

\bibitem{WolfWu08}
S. Wolf and J. Wullschleger.
\newblock ``New monotones and lower bounds in unconditional two-party
  computation,''
\newblock {\em IEEE Transactions on Information Theory}, vol. 54, no. 6, pp. 2792--2797,
  2008.

\bibitem{Wullschleger08thesis}
J. Wullschleger.
\newblock Oblivious-Transfer Amplification.
\newblock Ph.D. thesis, Swiss Federal Institute of Technology, Z\"{u}rich, 2008.
\newblock \url{http://arxiv.org/abs/cs.CR/0608076}.

\bibitem{Wyner75}
A. D. Wyner, 
\newblock ``The common information of two dependent random variables,''
\newblock {\em IEEE Transactions on Information Theory}, vol. 21, no. 2, pp. 163--179,
 1975.

\bibitem{WynerZi73}
A. D. Wyner and J. Ziv,
\newblock ``Rate-distortion function for source coding with side
information at the decoder,''
\newblock {\em IEEE Transactions on Information Theory}, vol. 22, no. 1, pp. 1--11,
 1976.

\bibitem{Yamamoto94}
H. Yamamoto,
``Coding theorems for Shannon's cipher system with correlated source
outputs, and common information,''
\newblock {\em IEEE Transactions on Information Theory}, vol. 40, no. 1, pp. 85--95, 1994. 

\end{thebibliography}
\end{document}